\documentclass[twocolumn]{revtex4}          
\usepackage{multirow}
\usepackage{graphicx}
\usepackage{bm}
\usepackage{color}
\usepackage{amsmath}
\usepackage{amssymb}

\begin{document}

\title{Gyrotactic phytoplankton in laminar and turbulent flows: a dynamical systems approach}

\author{Massimo Cencini}
\affiliation{Istituto  dei  Sistemi  Complessi,  CNR,  via  dei  Taurini  19,  00185  Rome,  Italy  and  INFN  Tor  Vergata}
\author{Guido Boffetta}
\affiliation{Dipartimento di Fisica and INFN, Universit\`a di Torino, via P. Giuria 1, 10125 Torino, Italy}
\author{Matteo Borgnino}
\affiliation{Dipartimento di Fisica and INFN, Universit\`a di Torino, via P. Giuria 1, 10125 Torino, Italy}
\author{Filippo De Lillo}
\affiliation{Dipartimento di Fisica and INFN, Universit\`a di Torino, via P. Giuria 1, 10125 Torino, Italy}

\begin{abstract}
Gyrotactic algae are bottom heavy, motile cells whose swimming
direction is determined by a balance between a buoyancy torque
directing them upwards and fluid velocity gradients.  Gyrotaxis has, in recent
years, become a paradigmatic model for phytoplankton motility in
flows.  The essential attractiveness of this peculiar form of motility
is the availability of a mechanistic description which, despite its
simplicity, revealed predictive, rich in phenomenology, easily
complemented to include the effects of shape, feed-back on the fluid
and stochasticity (e.g. in cell orientation). In this review we
consider recent theoretical, numerical and experimental results to
discuss how, depending on flow properties, gyrotaxis can produce
inhomogeneous phytoplankton distributions on a wide range of scales,
from millimeters to kilometers, in both laminar and turbulent flows.
In particular, we focus on the phenomenon of \textit{gyrotactic
  trapping} in nonlinear shear flows and in fractal clustering in
turbulent flows. We shall demonstrate the usefulness of ideas and tools
borrowed from dynamical systems theory in explaining and interpreting
these phenomena.

\keywords{Microorganisms in a flow \and Fractal clustering \and Motility}
\end{abstract}

\maketitle
\section{Introduction}
\label{sec1}

Biological and geophysical fluids host a sea of microorganisms many of
which are motile. An often overlooked aspect of the life of such
microorganisms is that the fluids where they are suspended are not
still but flowing. For instance, in fresh water and marine environments
microorganisms, such as unicellular algae or bacteria, are exposed to
turbulent motion \cite{guasto2012fluid} and turbulence is believed to
have been one of the the main factors in shaping the huge variety of form and
strategies of such aquatic microbes \cite{margalef1978life}. In
(photo-)bio-reactors microorganisms are grown in continuously stirred
tanks \cite{rodolfi2009microalgae,croze2013dispersion}. Bacteria
composing the microbioma of mammals undergo flowing fluids in several
organs that they inhabit such as, e.g., the gut or renal tissues
\cite{duan2008shear,kim2012human}. Microbes are also exposed to a
variety of water flows in the soil \cite{ebrahimi2014microbial}.

In the presence of flows, microorganisms are at the mercy of the
velocity field which transports them possibly, for motile species, in
addition to their intrinsic swimming, and modifies their orientation
by action of velocity gradients. Flow can thus impact the motility of
microorganism, their spatial distribution, interaction with surfaces,
response to nutrients or other chemicals for motile species
\cite{rusconi2015microbes,chilukuri2014impact,locsei2009run,stocker2012ecology},
and nutrient uptake also for non motile ones
\cite{karp1996nutrient}. Moreover, flow influences the encounter rate
of microorganisms \cite{rothschild1988small}, affecting their
reproduction and competition
\cite{kiorboe1995turbulence,pigolotti2013growth}. In summary, flows are
key in shaping microbial responses and ecology
\cite{kiorboe2008mechanistic}.

In this brief review we aim at emphasizing some interesting phenomena
that can emerge due to the modification of the microorganisms'
swimming direction by velocity gradients, which affects both the
individual motion of microorganisms and their spatial distribution in
dilute suspensions.  More specifically, we focus on the case study of
gyrotactic phytoplankton.

Phytoplankton is responsible for about half of the photosynthetic
activity on Earth. It is composed by thousands of species many of
which are able of swim. Motility confers phytoplankton the ability to
reach well-lit waters near the sea surface during daylight and migrate
into deeper water, richer of nutrients, during the night
\cite{lampert1989adaptive}.  For several species, the upward migration
is guided by a stabilizing torque, induced, e.g., by bottom heaviness,
which biases cell’s swimming direction upwards, and is opposed by
hydrodynamic shear, which exerts a viscous torque tending to overturn
the cell.  When the swimming direction results from the competition
between the cell’s stabilizing torque and the shear-induced viscous
torque, we speak of \textit{gyrotaxis}
\cite{kessler1985hydrodynamic,Pedley1987,Pedley1992}.  In both laminar and
turbulent flows gyrotaxis can promote heterogeneous spatial
distribution of phytoplankton cells.

Unicellular algae of the genus \textit{Chlamydomonas} aggregate in the
center (walls) of downwelling (upwelling) vertical pipe flows
\cite{kessler1985hydrodynamic}, a similar phenomenon can be induced by
phototaxis in horizontal pipe flows
\cite{garcia2013light,martin2016photofocusing}. Gyrotaxis not only alters cells' spatial
distribution but also the cells' dispersion properties
\cite{thorn2010transport,bees2010dispersion,bearon2012biased}, which can be
important for photobioreactors \cite{croze2013dispersion}.  When the
fluid acceleration is not negligible with respect to the gravitational
one, the stabilizing torque biases the motion in a position dependent
direction given by superposition of fluid and gravitational
acceleration. This causes, for instance, cell focusing toward the rotation
axis, when this is directed along the vertical
\cite{delillo2014turbulent,cencini2016centripetal}.  Remarkably, also
when the shear-induced viscous torque is not balanced by the
stabilizing one, interesting phenomena can happen. In this condition,
cells overturn due to the shear induced torque and start tumbling
without directed motion.  In inhomogeneous shear flows, this tumbling
motion can trap cells in regions of high shear, a phenomenon,
discovered in microfluidic experiments \cite{durham2009disruption},
which can explain the formation of high phytoplankton concentrations
in very thin layers as observed in coastal oceans
\cite{durham2012thin}. This gyrotactic trapping has an interesting
interpretation from a dynamical systems point of view
\cite{Santamaria2014}. In more complex laminar flows, such as
Tayolor-Green vortices, the combination of gyrotactic motility and
flow can give rise to small scale aggregation and complex trajectories
\cite{durham2011gyrotaxis}.  Such effects become even more striking in
turbulent flows of moderate intensity, where gyrotaxis can generate
small-scale fractal clusters which are dynamically formed and
dissolved, both when the fluid acceleration can be neglected and when
it cannot \cite{durham2013turbulence,delillo2014turbulent}, see also
\cite{fouxon2015phytoplankton,gustavsson2016preferential}. The
properties of such fractal clusters is also influenced by cell
morphology \cite{gustavsson2016preferential,zhan2014accumulation}. The physics underlying the formation of such fractal clusters can be easily
understood using dynamical systems concepts: cells swimming in turbulence can be described in terms of a chaotic, dissipative system, whose trajectories naturally evolve onto (multi~-~)fractal sets
\cite{ott1993}.

Here, we review some of the above phenomena emphasizing their
interpretation within the framework of dynamical systems theory. This
point of view is chosen on the basis of the background of the Authors.
This might have biased the choice of some topics and it is inevitable that some
relevant works on the subject have been not properly
discussed. This choice also left out many interesting phenomena
arising in dense suspensions of gyrotactic organisms such as
bioconvection \cite{pedley1988growth,ghorai2007gyrotactic} or complex
rheological effects \cite{rafai2010effective}. Before presenting the
organization of the matter, we would like to mention some interesting
phenomena induced by the interplay between motility and flow that
arise in other kinds of microorganisms, such as bacteria, and that are
connected with those observed in gyrotactic phytoplankton.

For instance, owing to their elongated shape, bacterial cells swimming
in a pipe (Poiseuille) flow are affected by Jeffery orbits
\cite{jeffery1922motion} which can induce upstream swimming in low
shear regions and tumbling in high shear ones. This was predicted on
the basis of mathematical models in \cite{Stark_PRL2012,Stark_EPJE2013}
and observed in microfluidic experiments \cite{rusconi2014bacterial},
which revealed also the accumulation of cells in high shear regions
due to trapping induced by Jeffery orbits. This trapping causes
spatial inhomogeneity and can reduce the efficiency of chemotaxis
\cite{rusconi2014bacterial}, as also confirmed by mathematical
analysis \cite{bearon2015trapping}. Motility also modifies the
transport properties with respect to passively advected particles
\cite{khurana2011reduced}.  More in general, flow effects on motility
are expected to impact chemotaxis and other kinds of taxis
\cite{stocker2012ecology}.  Although such effects are largely unexplored,
we mention here the numerical study of chemotaxis in the presence of
a turbulent flow \cite{taylor2012trade}.
The details of cell morphology, such as the chirality of
the flagella, can also induce further directional biases in the
presence of velocity gradients \cite{Marcos2012}. Flows can also
impact the interaction of microorganisms with surfaces
\cite{chilukuri2014impact}, e.g., inducing accumulation
\cite{berke2008hydrodynamic} and upstream swimming
\cite{kaya2012direct} close to the walls of pipe flows. Fractal
clustering, similar to that found in gyrotactic cells, was also found
in models of bacteria swimming in cellular and chaotic flows with or
without taxis \cite{Torney2007,Torney2008}.

The material is organized as follows. In Sect.~\ref{sec2} we describe 
the equations which have become the standard model of
gyrotactic motility and their modification when fluid acceleration
is important. Moreover, we briefly discuss the experimental validation
of that modification and the generic phenomenology one can derive from these
models. Section~\ref{sec:kolmogorov} is devoted to the phenomenon of
gyrotactic trapping in inhomogeneous shear flows. In particular, we
will consider the case of the Kolmogorov flow and show how, borrowing
ideas from conservative dynamical systems, gyrotactic trapping can be
ascribed to the presence of effective barriers to transport. We will
also briefly discuss how the presence of small-scale turbulence
destabilizes the trapping. In Sect.~\ref{sec:turbo}, we shall focus on
the effects of turbulence on gyrotactic motility, showing how fractal
clustering emerges due to the chaotic dissipative character of the
dynamics.  Moreover, we will link this phenomenon to the well known
clustering of inertial particles in turbulence, showing that there are
many conceptual analogies. We will also briefly consider the case of a
population in which cells have different characteristics, closer to what one
may observe in the ocean, to discuss how clustering can be revealed in these cases.
Finally, Sect.~\ref{sec:conclusions} is devoted to conclusions.

\section{Mathematical models\label{sec2}}

Gyrotaxis is observed in several species of unicellular algae such as,
e.g., biflagellate spheroidal algae belonging to the genus
\textit{Chlamydomonas} and \textit{Dunaliella} , and
also in some monoflagellate species such as e.g. \textit{Heterosigma
  akashiwo} \cite{kessler1985hydrodynamic,guasto2010oscillatory,durham2009disruption,durham2013turbulence}. Having as a reference \textit{Chlamydomonas}, we
can consider the cell close to spherical with a diameter of about $10\mu{\rm
  m}$ and swimming speeds around $100\mu{\rm m}\, s^{-1}$ 
\cite{guasto2010oscillatory}. Most 
phytoplankton cells have density very close to that of water and thus
can be considered as neutrally buoyant. A characteristic of most cells
displaying gyrotaxis is to have an inhomogeneous distribution of mass
leading to a displacement of the cell center of gravity with respect
to its center of symmetry. In particular, their center of mass is displaced opposite to their direction of swimming, so they are generally defined as bottom-heavy
\cite{kessler1985hydrodynamic}. 

The mathematical model for gyrotactic algae was introduced by Kessler
\cite{kessler1985hydrodynamic} (see also
\cite{pedley1990new,Pedley1992}) on the basis of the observation that
bottom-heavy swimming micro-organisms focus in the center of a pipe
when the fluid flows downwards. The swimming direction ${\bf p}$ results
from the competition between gravity-buoyant torque, due to
bottom-heaviness, and the shear-induced viscous torque and evolves
according to

\begin{eqnarray}
\nonumber \dot{\bf p} =\frac{1}{2B} \left[\hat{\bf z}-(\hat{\bf z} \cdot {\bf p}) \bf p \right] + \frac{1}{2} {\bm \omega} \times {\bf p}&+& \alpha [\hat{S}\mathrm{\bf p}-(\mathrm{\bf p}\cdot \hat{S}\mathrm{\bf p})\mathrm{\bf p}]\\
&+&{\bf \Gamma}_r
\label{eq2.1}
\end{eqnarray}
where ${\bm \omega}={\bm \nabla} \times {\bf u}$ is the vorticity at
the position of the cell, $B=\nu \alpha_\perp/(2h g)$ is a characteristic
orientation time which depends on viscosity $\nu$, gravity ${\bf g}=-g
\hat{\bf z}$, on the displacement $h$ of the cell center of mass
relative to the geometrical center and on $\alpha_\perp$ the dimensionless
resistance coefficient for rotation about an axis perpendicular to $\bf p$.
For a sphere $\alpha_\perp=6$ and thus $B=3 \nu/(h g)$.
The third term on the rhs is the rotation due to local rate of strain $\hat{S}_{ij}= \frac12
(\partial_ju_i+\partial_iu_j)$ and it is controlled by the shape factor $\alpha=(l^2-d^2)/(l^2+d^2)$ measuring the elongation $l$ with respect to the width
$d$ of the cell \cite{jeffery1922motion}. Prolate (respectively oblate) particles have $\alpha>0$ (respectively $\alpha<0$) and the particular cases of rods, disks and spheres are described by $\alpha=1,-1,0$. As a consequence, this term vanishes for spheres. The stochastic term ${\bf \Gamma}_r$ represents rotational diffusion of the swimming direction
as a results of irregularities in the cell propulsion
\cite{hill1997biased}, indeed cells are too large to be subjected to
Brownian rotation.  Equation~(\ref{eq2.1}) has been written assuming
the general case of ellipsoidal cells, although in this review we will mainly
consider spherical cells, and we will thus assume $\alpha=0$ unless otherwise specified.

Owing to their small size and small density mismatch with the fluid,
gyrotactic cells can be represented as point-like and
neutrally buoyant particles transported by the fluid velocity ${\bf
  u}({\bf x},t)$ with a superimposed swimming velocity of intensity
$v_s$ along the direction ${\bf p}$
\begin{equation}
\dot{\bf x}= {\bf u} + v_s {\bf p}\,.
\label{eq2.2}
\end{equation}

We remark that in writing the model defined by
Eqs.~(\ref{eq2.1})-(\ref{eq2.2}) many details have been neglected
including the unsteadiness of swimming due to flagella beating,
cell-–cell interactions, the feedback of cell motion on the
surrounding fluid, and (see below) effects due to fluid
acceleration. The model can thus be appropriate only for dilute
suspensions where interactions between cells and flow modifications
are expected to be negligible. In spite of its simplicity, however,
the gyrotactic model (\ref{eq2.1}-\ref{eq2.2}) is able to predict
remarkable features observed in experiments, such as the focusing
observed in pipe flows \cite{kessler1985hydrodynamic}.

To illustrate the phenomenon of focusing, we consider a
two-dimensional, laminar flow in a vertical pipe, in which the velocity
field has only the vertical component $w(x)=U (1-(x/L)^2)$ and the
coordinate $x \in [-L:L]$ varies in the wall-normal direction.
Neglecting the stochastic term, with this choice it is easy to find
the equilibrium, quasi stationary solution of Eq.~(\ref{eq2.1}):
\begin{equation}
\begin{array}{l}
\mathrm{p}^{eq}_x = 4 U B x/L^2 \\
\mathrm{p}^{eq}_z = \left(1 - 16 B^2 U^2 x^2/L^4 \right)^{1/2}
\end{array}
\label{eq2.2-1}
\end{equation}
The equilibrium solution exists for $B U/L \le 1/4$, i.e. for small
{\it stability number} $\Psi \equiv BU/L$, which is the ratio of the
second to the first term in the RHS of (\ref{eq2.1}).  When used in
(\ref{eq2.2}), the stationary solution (\ref{eq2.2-1}) shows that for
downwelling flow ($U<0$) the horizontal swimming direction is toward
the symmetry axis (since $\mathrm{p}_x \propto -x$). This leads to the
{\it gyrotactic focusing}, i.e. an accumulation of motile cells at the
center of the pipe \cite{kessler1985hydrodynamic}. For upwelling flows
($U>0$) the sign reverses and accumulation happens at the pipe's
walls. When the stochastic term is included one realizes that
cells focus in a region with a finite width controlled by the noise
strength \cite{pedley1990new,Pedley1992}.

In analogy to the stability numbers, from (\ref{eq2.2}) we also
introduce a {\it swimming number} defined as $\Phi=v_s/U$, measuring
the swimming speed with respect to the fluid velocity.  More complex
dynamics of gyrotactic swimmers in laminar flows, with or without
stochastic behavior, have been investigated \cite{thorn2010transport}.

\subsection{Effects of fluid acceleration}

Recently, the gyrotactic model has been generalized by taking into
account the effect of fluid acceleration in the orientation of the
cell \cite{delillo2014turbulent}.  By introducing the total
acceleration on the cell ${\bf A}={\bf g}-{\bf a}$, where ${\bf
  a}=d{\bf u}/dt$, Eq.~(\ref{eq2.1}) becomes
\begin{equation}
\dot{\bf p}= - {1 \over v_0} \left[{\bf A}-({\bf A} \cdot {\bf p}) \bf p
\right] + {1 \over 2} {\bm \omega} \times {\bf p} + {\bf \Gamma}_r
\label{eq2.3}
\end{equation}
where $v_0=B g$ is the reorientation speed.

The general model (\ref{eq2.3}) has been validated by laboratory
experiments of a suspension of gyrotactic cells in a cylindrical
vessel of radius $R$ which rotates with constant angular velocity $\Omega$
\cite{cencini2016centripetal}. When rotation is sufficiently fast,
centripetal acceleration overcomes gravity and the cells are expected
to swim towards the axis of the cylinder.
Neglecting the stochastic term, the deterministic motion given by
(\ref{eq2.1}-\ref{eq2.2}) in the velocity field produced by solid
body rotation ${\bf u}=(-\Omega y, \Omega x,0)$ can be easily obtained
in cylindrical coordinates ${\bf x}=({\bf r},z)$
under the hypothesis of local equilibrium in the
swimming direction, i.e. by assuming $\dot{\bf p}=0$ locally.
In this limit one obtains \cite{cencini2016centripetal}
\begin{equation}
{\bf p}^{eq}=\left({-\gamma {\bf r} \over \sqrt{1+(\gamma r)^2}},
{1 \over \sqrt{1+(\gamma r)^2}} \right)
\label{eq2.4}
\end{equation}
where $\gamma=\Omega^2/g$.
For $\gamma r \ll 1$, from (\ref{eq2.2}) and (\ref{eq2.4}) we have
$\dot{r}=- \gamma v_s r$ which implies that the cell position relaxes
exponentially towards the rotation axis $r=0$ as $r(t)=r(0) e^{-\gamma v_s t}$.

\begin{figure}[htb!]
\begin{minipage}[c]{0.4\columnwidth}
\includegraphics[width=\columnwidth]{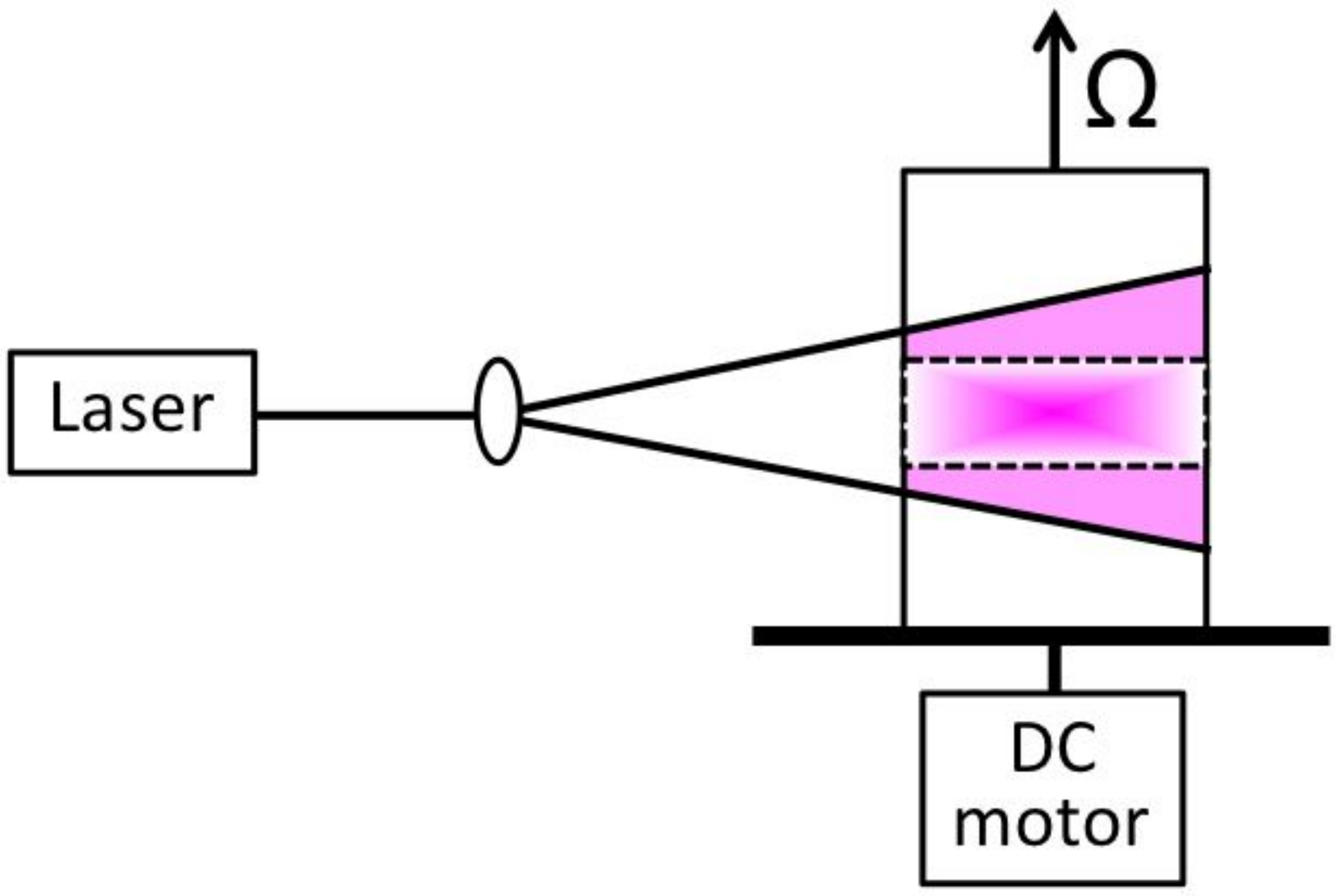}
\end{minipage}
\begin{minipage}[c]{0.5\columnwidth}
\includegraphics[width=\columnwidth]{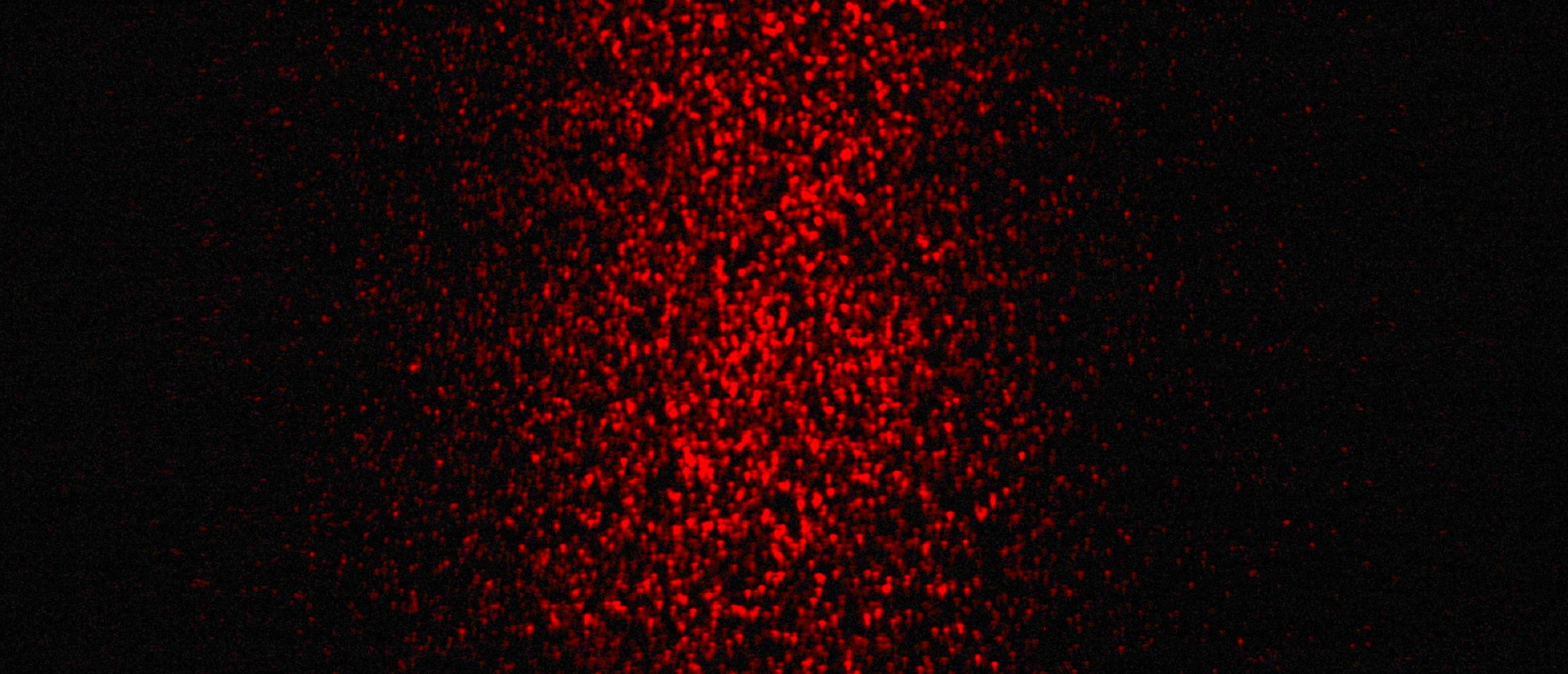}
\vspace{2mm}
\includegraphics[width=\columnwidth]{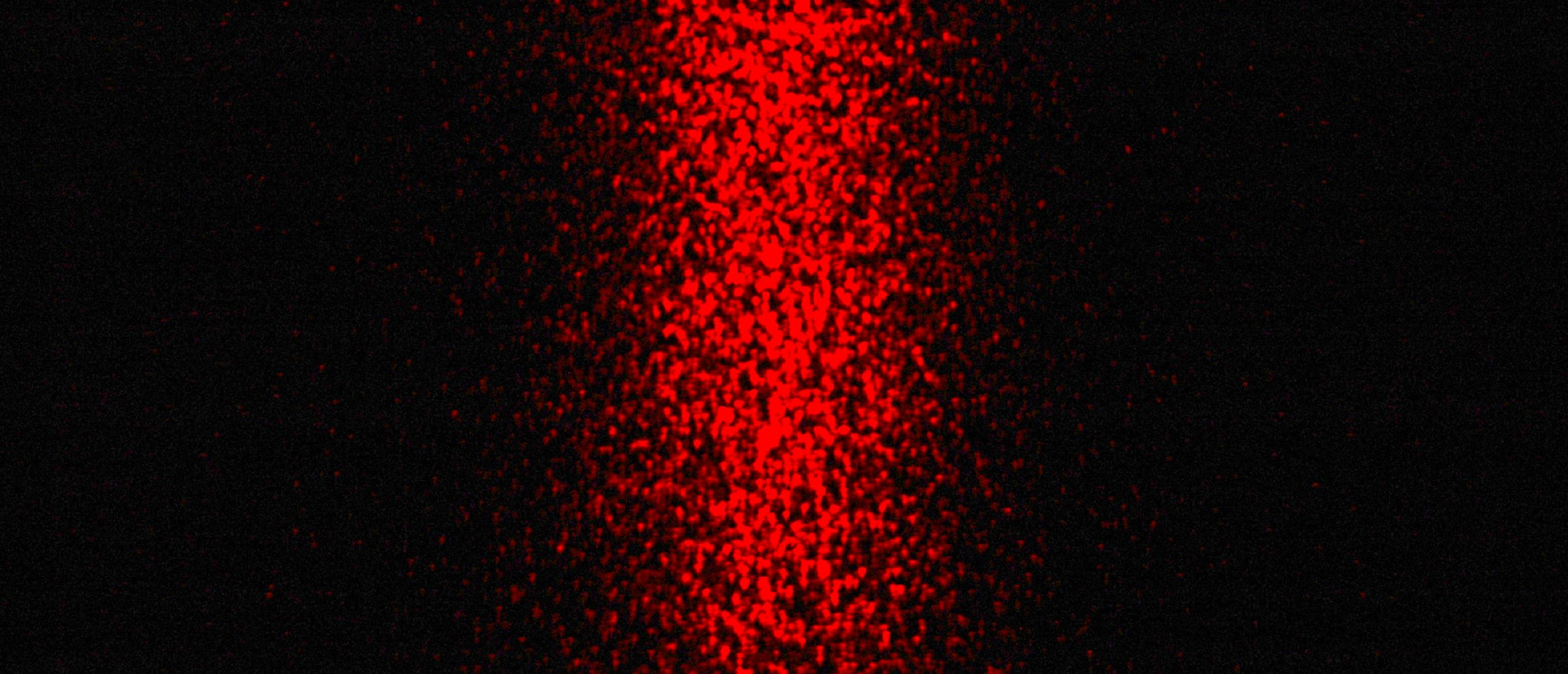}
\end{minipage}
\caption{
Experimental setup for the validation of the model (\ref{eq2.3}).
A cylindrical vessel is filled with a suspension of {\it C. augustae}
and placed over a table rotating with constant angular velocity
$\Omega=2 \pi f$.  A blue laser (wavelength $450 \, nm$)
is used to induce fluorescence in
the cells and their images are taken by a CCD camera at resolution
$3000 \times 2000$ pixels with a low-pass red filter at $600 \, nm$.
The two pictures on the right are examples of the images (central part)
taken by the camera at the final time $t=600\, s$ for $f=5\, Hz$ (top) and
$f=8\, Hz$ (bottom).
}
\label{fig1}
\end{figure}

Figure~\ref{fig1} shows a sketch of the experimental setup together
with two pictures of the asymptotic cell distributions for different
angular velocities.  The experiments show that, at late times, the
asymptotic cell distribution is over a finite radius around the axis,
and this has been interpreted as due to stochastic effects, modeled as
rotational diffusion, which deviate the swimming direction with
respect to the equilibrium one (\ref{eq2.4}).

\begin{figure}[htb!]
\includegraphics[width=\columnwidth]{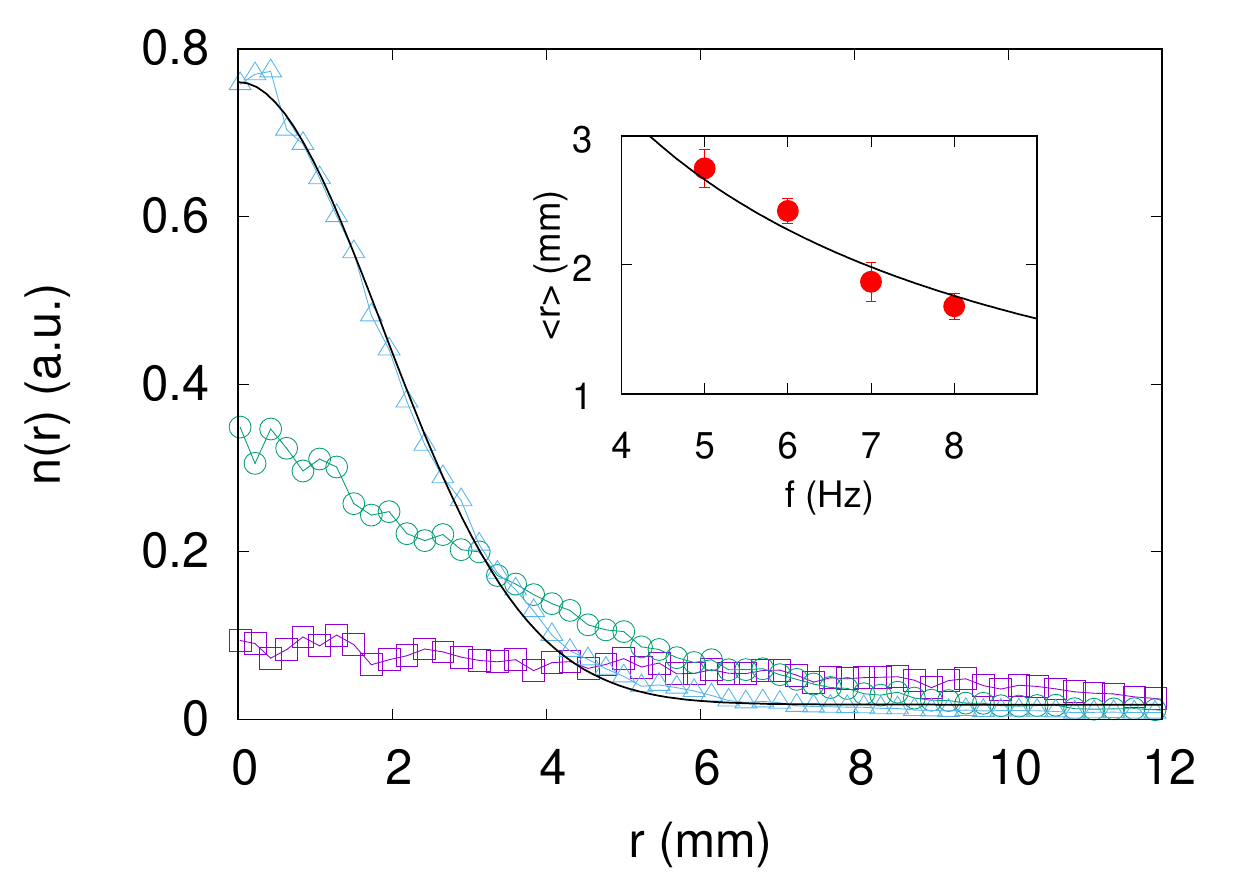}
\caption{
Evolution of the experimental radial population density
$n_{exp}(r,t)$ (in arbitrary units) for the experiment at frequency
$f=7\,Hz$ a time $t=150 \, s$ (pink squares), $t=300 \, s$ (green circles) and
at time $t=600 \, s$ (blue triangles) as a function of the distance
from the cylinder axis $r$. The continuous line is the theoretical prediction (\ref{eq2.5}).
Inset:
Average radius of the cell population, $\langle r \rangle$ in
asymptotic stationary conditions as a function of rotation frequency.
The line represents the theoretical prediction obtained from
(\ref{eq2.5}).
}
\label{fig2}
\end{figure}

The stationary distribution in presence of rotational diffusivity can
be obtained within the so-called Generalized Taylor Dispersion theory
\cite{frankel1989foundation,bearon2011spatial,bearon2012biased}.  The
basic idea is to reduce the Fokker-Planck equation for the probability
${\cal P}({\bf x},{\bf p},t)$ to an advection-diffusion equation for
the population density $n({\bf x},t)=\int d{\bf p} {\cal P}$ which
includes an effective drift and diffusivity tensor which can be
analytically derived in the approximation of fast orientation of the
swimming direction \cite{bearon2011spatial,bearon2012biased}. The
final result (see \cite{cencini2016centripetal} for details) is, for
$\gamma r \ll 1$, a Gaussian distribution of the radial population
density
\begin{equation}
n_s(r) = \mathcal{N} \exp\left(- {\gamma r^2 \over 2 v_s B F_3^2(\lambda)}
\right)
\label{eq2.5}
\end{equation}
where $F_3(\lambda)$ a dimensionless function of the parameter
$\lambda=1/(2 B d_r)$ ($d_r$ is the rotational diffusivity), which can
be analytically expressed as series, and the coefficient $\mathcal{N}$
can be written in terms of the total number of cells $N_s=\int dr
n_s(r)$ as $\mathcal{N}=N_s \gamma/(2 \pi H v_s B F_3^2(\lambda))$
($H$ is the vertical size of the vessel). Figure~\ref{fig2} shows
the evolution of the experimental radial population density
measured at different times, together with the asymptotic theoretical density
Eq.(\ref{eq2.5}) for the case $\gamma=0.20\,mm^{-1}$ with parameters
$v_s=0.1 \, mm$, $d_r=0.067 \, rad s^{-1}$
taken from the literature and $B=7.5 \, s$ as fitting parameter. 
   The inset of the
figure shows the average radius at stationarity as a function of the
rotation angular velocity. As one can see the full line, obtained on
the basis of Eq.(\ref{eq2.5}) (once the effect of background nonmotile
impurities is taken into account, see \cite{cencini2016centripetal}
for details), is in very good agreement with the experimental data
demonstrating the validity of the model (\ref{eq2.3}).

\begin{figure*}[th!]
\includegraphics[width=1\textwidth]{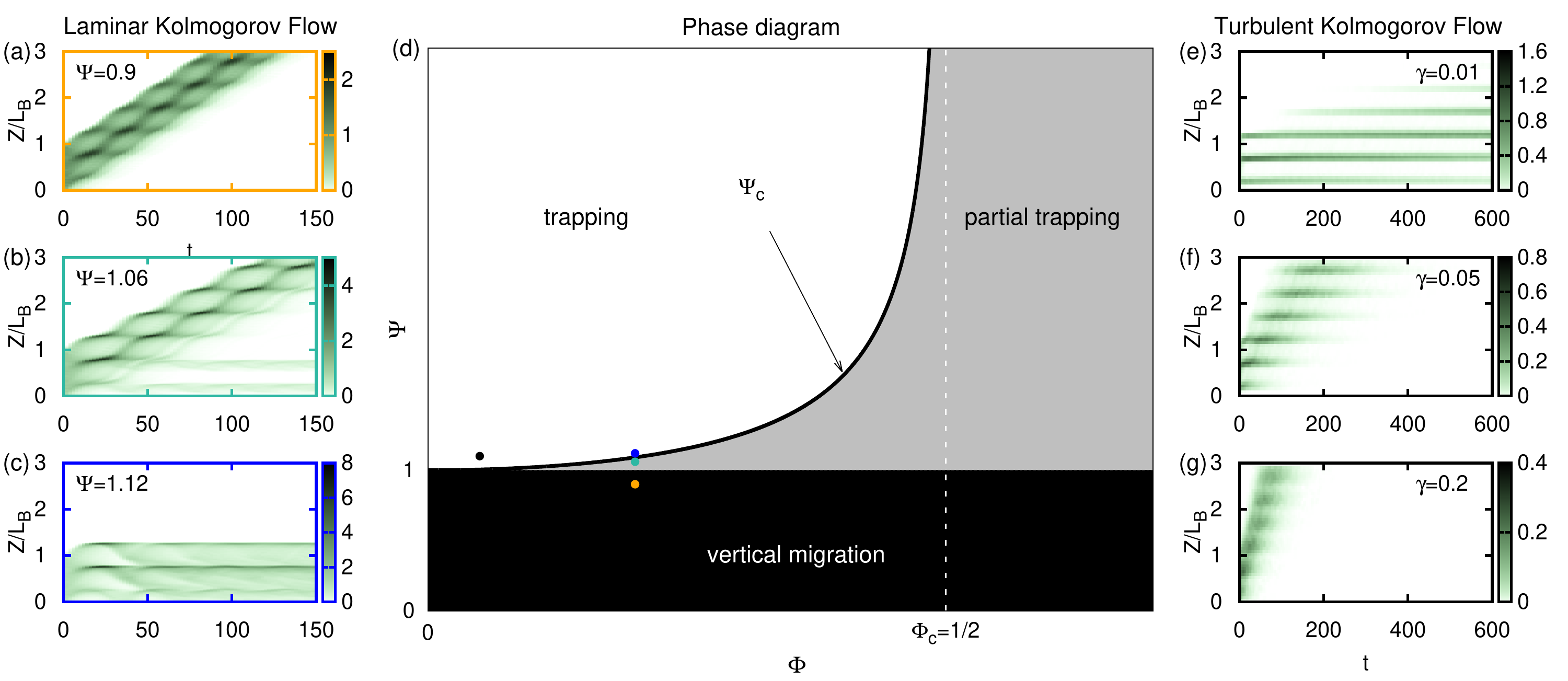}
\caption{(color online) Phenomenology of the gyrotactic swimming in
  the Kolmogorov flow. (a-c) Evolution of the vertical density of
  cells, $\rho(Z,t)$, in the 2d laminar Kolmogorov flow for $\Phi=0.2$
  with (a) $\Psi=0.9<\Psi_c$, (b) $1<\Psi=1.06<\Psi_c$ and
  $\Psi=1.12>\Psi_c$, with $\Psi_c=(1-4\Phi^2)^{-1/2}$. The density
  has been obtained coarse-graining the vertical position of $N=10^4$
  cells initialized uniformly in $(\theta,Z)\in
  [-\pi,\pi]\times[0:2\pi]$ and evolved by integrating
  Eqs.~(\ref{eq:2thdot}) and (\ref{eq:2zdot}) with a 4$^{th}$-order
  Runge-Kutta scheme. (d) Swimming behavior in parameter space
  ($\Phi,\Psi$) for the laminar Kolmogorov flow.  The white region
  corresponds to vertically trapped orbits ($\Psi>\Psi_c$), the grey
  one to partially trapped trajectories ($1<\Psi< \Psi_c$, with
  coexistence of trapped and vertically migrating cells) and, finally,
  the black to vertically migrating cells $(\Psi<1)$. The colored
  circles denote the parameters used in panels (a-c), whose border has
  the same color.  The black dot corresponds to the $(\Phi,\Psi)$
  values used in panels (e-g), which are the same as (a-c) but for the
  3d turbulent Kolmogorov flow for $\Phi=0.05$, in the trapping
  regime, $\Psi=1.1>\Psi_c$, at varying the intensity of turbulent
  fluctuations $\gamma=0.01$ (a), $0.05$ (b), $0.2$ (c). Time and
  scale have been made non-dimensional as for the laminar case. The
  thin layers are now transient and their duration decreases at
  increasing the intensity of the turbulent fluctuations (see
  Sect.~\ref{sec:turboKolmo} for a discussion).}
\label{fig3}
\end{figure*}

\section{Gyrotactic trapping in laminar and turbulent Kolmogorov flow\label{sec:kolmogorov}}


The mechanisms of cell accumulation described above rely on the assumption that
$\Psi< 1$, since they are based on the existence of a quasi-static solution of
Eq.(\ref{eq2.1}). However, when $\Psi>1$, the second term dominates and the cell tumbles with no persistent direction.

Laboratory experiments \cite{durham2009disruption} demonstrated that in
an inhomogeneous shear flow with vorticity varying along the gravity
direction, gyrotactic cells can get trapped where $\Psi>1$ locally.
Such \textit{gyrotactic trapping} has been proposed as a possible
mechanism for the formation of thin phytoplankton layers (TPLs) often
observed in ocean coastal areas
\cite{Dekshenieks2001,cheriton2007effects,churnside2009thin,Steinbuck2009}.  TPLs are
regions of high vertical concentration of phytoplankton, centimeters
to one meter thick, which extend horizontally up to kilometers and
last from hours to a few days (see \cite{durham2012thin} for a
review). Though TPLs can be formed by non-motile or motile species
characterized by different swimming styles, \textit{Heterosigma
  Akashiwo} -- a toxic, gyrotactic algae -- is known to form harmful
thin layers \cite{ryan2011harmful} and gyrotactic trapping could be an
effective explanation.  For the sake of completeness, we mention that
other mechanisms, not discussed here, of phytoplankton cells
accumulation in shear flows can emerge when the cell shape is
elongated, see e.g. \cite{barry2015}.


\subsection{The Kolmogorov Flow (KF)}
We focus on the Kolmogorov Flow (KF), a periodic shear flow
model, much studied for the transition to turbulence
\cite{Sivashinsky1985,She1987,Borue1996}, characterized by an
inhomogeneous distribution of vorticity, necessary for gyrotactic
trapping.  The KF solves the incompressible
Navier-Stokes equations subjected to a monochromatic body force:
\begin{equation}
\partial_t \bm u +\bm u \cdot \bm \nabla \bm u = - \bm \nabla p +\nu \Delta \bm u + F\cos(z/L) \hat{\bm x}\, 
\label{eq:ns}
\end{equation}
where $p$ is the pressure, density is taken to unity $\rho=1$, and
$\hat{\bm x}$ denotes the unit vector in the horizontal direction.
The physical domain is a cube of side $L_B=2\pi L$ with periodic
boundary conditions in all directions.  Equation (\ref{eq:ns}) admits
a stationary solution, the {\it laminar} KF $\bm u= U \cos(z/L)
\hat{\bm x}$ with $U= L^2 F/\nu$, which becomes unstable to large scale
perturbations when the
Reynolds number, $Re=UL/\nu$, exceeds the critical value
$Re_c=\sqrt{2}$.  The first instability is two-dimensional, at
increasing $Re$ three-dimensional motion develops and the flow
eventually becomes turbulent \cite{Borue1996,Musacchio2014}. Also in
the turbulent KF the time averaged velocity $\bar{\bm u}$
remains monochromatic, $\overline{\bm u}=U \cos(z/L) \hat{\bm x}$, but
with a different amplitude $U<L^2 F/\nu$ \cite{Musacchio2014}. This
latter property is useful to study the effect of turbulent
fluctuations, which will be briefly discussed at the end of the
section.

\subsection{Gyrotactic Cells in Laminar Kolmogorov Flow}

After non-dimensionalization by measuring lengths, velocities and
times in terms of $L$, $U$ and $L/U$, for the KF Eq.~(\ref{eq2.2})
reads
\begin{eqnarray}
\dot{X} &=& \cos{Z} + \Phi \mathrm{p}_x \label{eq:xdot}\\
\dot{Y} &=& \Phi \mathrm{p}_y \label{eq:ydot}\\
\dot{Z} &=& \Phi \mathrm{p}_z \label{eq:zdot}\,,
\end{eqnarray}
where $\Phi=v_s/U$ is the ratio of the swimming speed over the flow
velocity.  While, Eq.~(\ref{eq2.1}) becomes
\begin{eqnarray}
\dot{\mathrm{p}}_x &=& -\frac{1}{2\Psi} \mathrm{p}_x \mathrm{p}_z 
-\frac{1}{2} \sin Z\, \mathrm{p}_z 
\label{eq:pxdot}\\
\dot{\mathrm{p}}_y &=& -\frac{1}{2\Psi} \mathrm{p}_y \mathrm{p}_z 
\label{eq:pydot}\\
\dot{\mathrm{p}}_z &=& \frac{1}{2\Psi} (1-\mathrm{p}^2_z) 
+\frac{1}{2} \sin Z\, \mathrm{p}_x \,.
\label{eq:pzdot}
\end{eqnarray}
where $\Psi=B U/L$ measures the cells' stability. The box size in
dimensionless units is $L_B/L=2\pi$. Capital letters denote the
coordinates in non-dimensional units.

Considering that $|\mathrm{\bf p}|=1$, and noticing that $X$ and $Y$
do not enter the dynamics of the other variables, the dynamics is
effectively three dimensional. Moreover, using
Eqs.~(\ref{eq:zdot})-(\ref{eq:pydot}) and
Eqs.~(\ref{eq:zdot})-(\ref{eq:pxdot}) one can easily check that the
system admits the two following conserved quantities, respectively:
\begin{eqnarray}
  \mathcal{C}({\bf p},Z) &=& \mathrm{p}_y e^{Z/(2\Phi\Psi)} \label{eq:C1}\\
  \mathcal{H}({\bf p},Z)&=&\Phi e^{\frac{Z}{2\Phi\Psi}} \left[\mathrm{p}_x-\frac{\Psi (2\Phi\Psi\cos Z- \sin Z)}{1+4\Phi^2\Psi^2} \right]\,. \label{eq:C2}
\end{eqnarray}
Thus the dynamics has only one degree of freedom and it is integrable
(see \cite{Stark_PRL2012,Stark_EPJE2013} for a similar approach for
bacteria swimming in a shear flow).

For vertically migrating cells, the conservation of $\mathcal{C}$ in
Eq.~(\ref{eq:C1}) implies that $p_y$ should vanish exponentially as
$Z$ increases. This observation tells us that asymptotically the
motion is described by the two-dimensional dynamics
\begin{eqnarray}
\dot{\theta} &=& \frac{1}{2\Psi} \cos\theta +\frac{1}{2} \sin Z
\label{eq:2thdot}\\
\dot{Z} &=& \Phi \sin\theta \,,
\label{eq:2zdot}
\end{eqnarray}
where $(\mathrm{p}_x,\mathrm{p}_z)=(\cos\theta,\sin\theta)$.  Strictly
speaking the reduction to the above two dimensional system applies
only for vertically migrating cells, however the following derivations
extend also to the general three-dimensional dynamics also for
non-vertically migrating cells, see \cite{Santamaria2014} for details.
Here below we will denote with $Z$ the unrestricted vertical coordinate
and with $z=Z \mod 2\pi$ its restriction to the periodic cell, while
$\theta\in [-\pi,\pi]$.\footnote{We remark that $\mathcal{H}$
  (\ref{eq:C2}) is not periodic in $Z$: when $Z \to Z\pm 2 \pi\,n$, we
  have $\mathcal{H}(\theta,z) \to \mathcal{H}(\theta,z) e^{\pm \pi
    n/(\Phi\Psi)}$. We also notice that $\mathcal{H}(\theta,z)$ plays
  a similar role to the Hamiltonian and though the system is not
  Hamiltonian it can be made so by a non-canonical change of variables
  \cite{Santamaria2014}. }

For $\Psi<1$, Eqs. (\ref{eq:2thdot})-(\ref{eq:2zdot}) do not admit
fixed points and $Z$ grows in time.  The exponential dependence on $Z$
in (\ref{eq:C2}) and the conservation of $\mathcal{H}$ implies that
the term in square brackets in (\ref{eq:C2}) must decrease
exponentially with $Z$. Therefore, for large $Z$, the swimming angle
depends on $z$ (i.e. restricted on the torus) as given by
\begin{equation}
\cos\theta=\mathrm{p}_x=\frac{\Psi (2\Phi\Psi\cos z- \sin z)}{1+4\Phi^2\Psi^2}\,.
\label{eq:pxasym}
\end{equation}
Thus the vertical velocity will change with height and cells will
accumulate where it is minimal, i.e. for
$Z=n\,\pi-\arctan[1/(2\Phi\Psi)]$, for any integer $n$. Around these
positions one observes ephemeral layers of high density of cells, even
though cells are trapped asymptotically . This is confirmed in
Fig.~\ref{fig3}a showing the time evolution of the vertical
probability density distribution (PDF), $\rho(Z,t)$, resulting from an
initially uniform distribution in $Z \in [0:2\pi]$ for
$\Psi<1$. The transient accumulations last longer for smaller
values of $\Phi$.

Since $|\cos\theta|\leq 1$, after some algebra, one can realize that
Eq.~(\ref{eq:pxasym}) corresponds to a well defined relation between
$\theta$ and $z$ on the torus $[-\pi:\pi]\times[0:2\pi]$ either for
fast swimmers $\Phi\geq \Phi_c=1/2$ or if $\Phi<\Phi_c$ whenever
$\Psi<\Psi_c=(1-4\Phi^2)^{-1/2}$. Considering first the case
$\Phi<\Phi_c$, since $\Psi_c>1$ we have two possible instances.
For $1<\Psi<\Psi_c$ (Fig. \ref{fig3}b), we observe the coexistence
of upward migrating cells that, as for the $\Psi<1$ case,
asymptotically satisfy Eq.~(\ref{eq:pxasym}), and of cells remaining
vertically confined with high concentration in thin regions unrelated
to the minima of $\mathrm{p}_z$.  Conversely, for $\Psi>\Psi_c$
(Fig. \ref{fig3}c), we observe that all cells are asymptotically
vertically trapped and organized in thin layers.

\begin{figure}[t!]
\includegraphics[width=1\columnwidth]{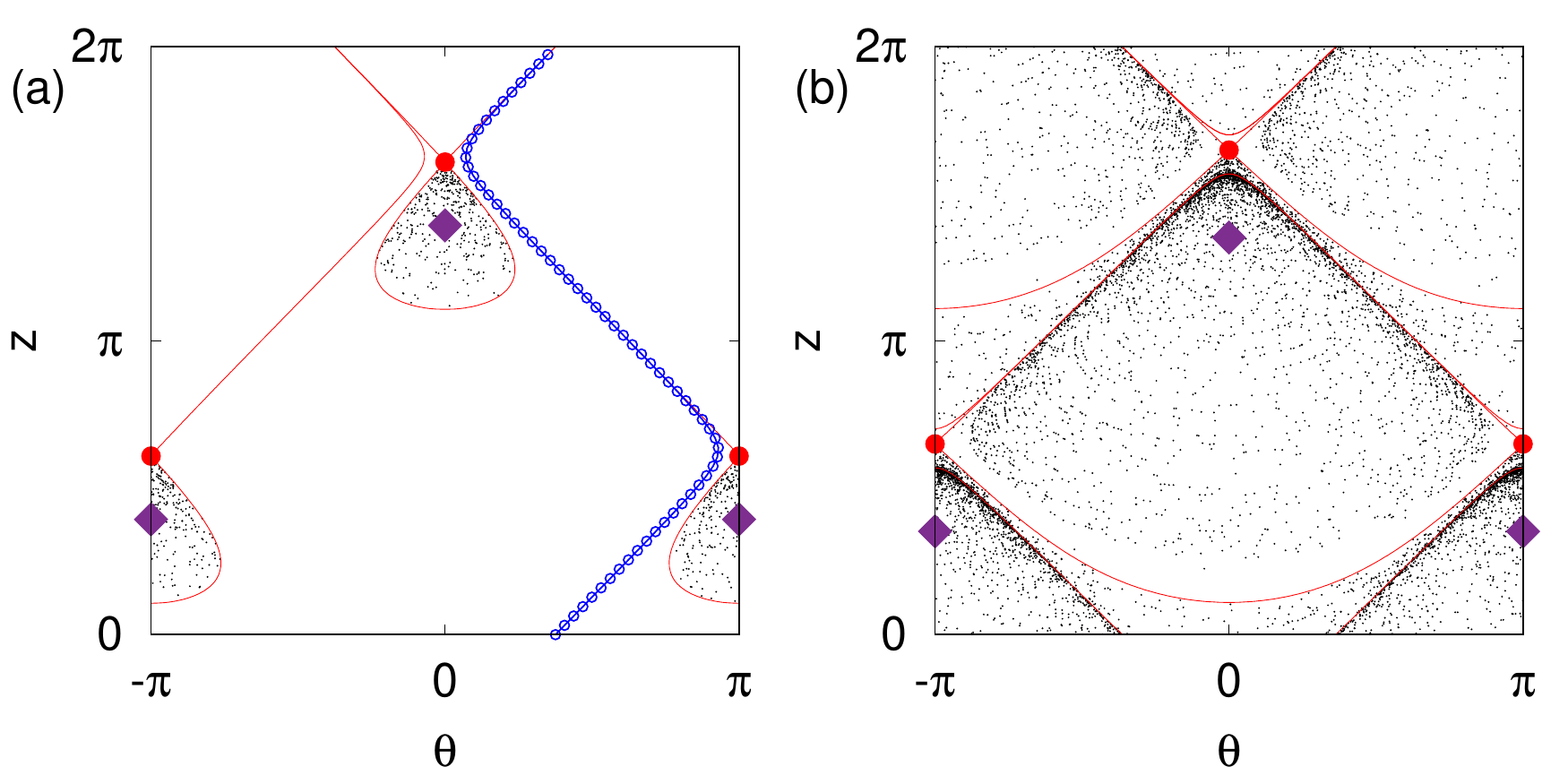}
\caption{Cells positions in the laminar 2D Kolmogorov flow at a long
  time on the $(\theta,z)$ torus for (a) $\Psi=1.06<\Psi_c$ and (b)
  $\Psi=1.12>\Psi_c$ with $\Phi=0.2$, corresponding to
  Fig.~\ref{fig3}b and c, respectively.  Red circles (purple
  diamonds) mark the hyperbolic (elliptic) fixed points. Red curves
  denote the separatrices emerging from the hyperbolic fixed points.
  At $\Psi<\Psi_c$ (a) small black dots corresponds to trapped orbits
  and empty blue circles to the orbit (\protect\ref{eq:pxasym}) which
  asymptotically characterized the vertically migrating cells and is
  very close to the separatrix.  For $\Psi>\Psi_c$ (b) all orbits are
  trapped, being confined between the separatrices.}
\label{fig4}
\end{figure}

The origin of these two behaviors can be traced back to the properties
of the phase-space dynamics on the $(\theta,z)$ torus. We start
noticing that for $\Psi>1$, Eqs. (\ref{eq:2thdot})-(\ref{eq:2zdot})
possess two hyperbolic (at $(\theta^*,z^*)=(0,2\pi-\arcsin\Psi^{-1})$
and $(\pi,\pi-\arcsin\Psi^{-1})$) and two elliptic fixed points (at
$(\theta^*,z^*)= (0,\pi+\arcsin\Psi^{-1})$ and
$(\pi,\arcsin\Psi^{-1})$).  However, depending on $\Psi$ being smaller
or larger than $\Psi_c$ the separatrices, corresponding to the orbits
emerging from the hyperbolic fixed points, qualitatively
change (see Fig.~\ref{fig4}). $1<\Psi<\Psi_c$
(Fig.~\ref{fig4}a), the hyperbolic point is the vertex of a
slip-knot containing the elliptic fixed point. Orbits whose initial condition
falls into this loop cannot escape it, those starting outside it
migrate vertically asymptotically following (\ref{eq:pxasym}). We thus
have only \textit{partial trapping}.  Conversely, for $\Psi>\Psi_c$
(Fig.~\ref{fig4}b), the separatrices roll up around the torus in
the $\theta$ direction becoming impenetrable barriers to vertical
transport and thus trapping the cells. As clear from
Fig.~\ref{fig4}, the compression between the elliptic and
hyperbolic points is responsible for the higher concentration layers
observed in Fig.~\ref{fig3}b,c.  When $\Phi>\Phi_c$ only purely
vertically migrating cells (for $\Psi<1$) or partial trapping (for
$\Psi>1$) is possible. Figure~\ref{fig3}d summarizes the various
regime in the parameter space $(\Phi,\Psi)$.

\subsection{Gyrotactic Cells in Turbulent Kolmogorov Flow\label{sec:turboKolmo}}
Fluctuations of velocity and/or vorticity, due to turbulence, allow 
a swimmer to escape from the homoclinic loops in the partial trapping
regime or from the separatrices in the trapping one
(Fig.~\ref{fig4}). As a result, layers are less intense and,
more importantly, they become transient. Below, we briefly discuss the
transient dynamics of such layers, for further details see
\cite{Santamaria2014}. Such a phenomenology is indeed observed in
field experiments with simultaneous measurement of biological and
physical properties, which have shown that while thin layers are
weakly affected by turbulence of moderate intensity, stronger
turbulence dissolves them \cite{wang2010evolution,Sullivan2010}. As discussed
in \cite{Santamaria2014} a similar effect can be induced, in the
laminar KF, by the presence of rotational diffusion, but this effect
is typically much smaller than that due to turbulence.

Increasing the Reynolds number the Kolmogorov flow becomes unstable
and eventually turbulence for large values of $Re$.
Nonetheless, since the (time-averaged) mean flow remain
monochromatic, velocity and vorticity fields can be still decomposed
in the time-averaged fields with superimposed fluctuations, i.e. $\bm
u^\prime$ and $\bm \omega^{\prime}$ as $\bm u =U\cos(z/L) \hat{\bm x}
+ \bm u^\prime({\bm x},t)$ and $\omega = -(U/L)\sin(z/L)\hat{\bm y} +
\bm \omega^\prime({\bm x},t)$.  However, even at relatively low $Re$,
fluctuations are non-negligible with respect to the mean flow, indeed
it has been found that $u^\prime_{\rm rms}/U\simeq
0.5$\cite{Musacchio2014}. Conversely, in real oceans fluctuations are
typically much smaller than the mean flow being depleted by, e.g.,
stratification \cite{Thorpe2007}.  For this reason, here, while
solving Eq.~(\ref{eq:ns}) by means of direct numerical simulations
(see \cite{Santamaria2014} for details), gyrotactic swimmers are
evolved by modulating the fluctuations $\bm u^\prime$ and $\bm
\omega^\prime$ with a multiplicative factor $\gamma<1$, so to control
their intensity.  Besides the possibility to control fluctuation
intensity, another advantage of such an approach is that the statistical
properties of the turbulent fluctuations do not change with $\gamma$
as they would by introducing stratification.

\begin{figure}[t!]
\includegraphics[width=1\columnwidth]{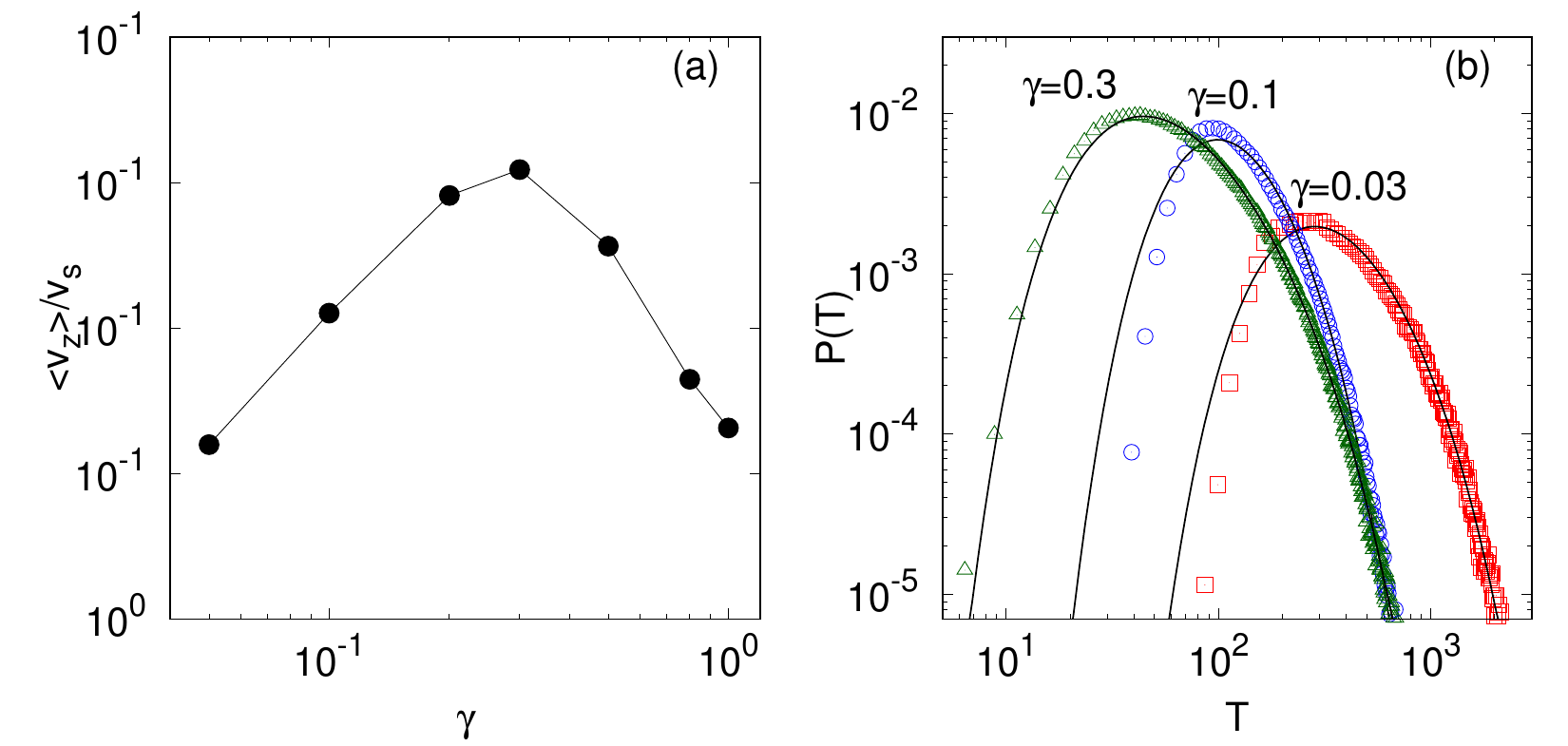}
\caption{Quantitative characterization of single cell and layer
  properties as a function of turbulent intensity $\gamma$, for
  $\Phi=0.05$ and $\Psi=1.1$. (a) Average vertical velocity $\langle
  v_z\rangle$ normalized to the swimming speed $v_s$.  (b) Exit time
  PDF for $\Psi=1.1$, and $\Phi=0.05$ at three turbulent intensities
  $\gamma$ as labeled, compared with the prediction (\ref{eq:ig}) with
  $V_d$ obtained from (a) and $D_z$ obtained as explained in the
  text. }
\label{fig5}
\end{figure}

Numerical simulations show that as soon as turbulent fluctuations are
considered (i.e. $\gamma>0$), even if $\Phi$ and $\Psi$ are chosen in
the trapping region (black circle in Fig.~\ref{fig3}d), vertical
migration becomes possible as clear by comparing
Figs.~\ref{fig3}e-g with Fig.~\ref{fig3}c.  At moderate
values of the turbulent intensity (Fig.~\ref{fig3}e), velocity
and vorticity fluctuations allow cells to escape from the trapping
regions by moving them to regions of lower shear, where upward
directed swimming is possible. Then cells get trapped around another
high shear region. We recall that the periodic layer structure is
inherited from the flow periodicity.  As a result, the average
vertical cell velocity, $\langle v_z \rangle$, which was zero in the
absence of turbulent fluctuations, becomes positive
(Fig.~\ref{fig5}a). However, very intense turbulence rotates the
cell swimming direction randomly and, moreover, fluctuations of the
vertical velocity also mix cells. As a consequence, the average
vertical motion $\langle v_z \rangle$ decreases for large values of
$\gamma$.  An intermediate turbulence intensity maximizes the vertical
migration velocity.

From the point of view of the single cell dynamics, the above
phenomenology means that, for low intensity fluctuations, a cell is
trapped for a finite time until fluctuations of the vertical velocity
or of the vorticity makes the cell able to escape its trapped
trajectory and to swim upwards until it gets trapped again. At high
turbulent intensity trapping is less and less effective, even
transiently, and the motion becomes basically diffusive in the
vertical direction due to turbulent diffusion.

The above observation suggests to model the vertical dynamics of
gyrotactic swimmers  in terms of a diffusive process with drift
$V_d$ (due to the average vertical migration speed, i.e. $V_d=\langle
v_z\rangle$) and diffusion constant $D_z$, both depending on turbulent
fluctuations. With this modelization in mind one can study the exit
times statistics -- a standard problem in stochastic processes
\cite{Redner} -- from a layer, i.e. of the times $T$ needed for a
swimmer to travel a vertical distance $L_B/2$, separating two
consecutive layers. For diffusion with drift, the
probability density function of the exit time $T$ is given by the
inverse Gaussian function, which  reads
\begin{equation}
\mathcal{P}(T)= \frac{L_B}{(4 \pi D_z T^3)^{1/2}} e^{-\frac{(V_d T -
    L_B/2)^2}{4D_z T}} \,.
\label{eq:ig}
\end{equation}
The above expression provides a prediction for the exit time PDF that
can be directly tested against the measured one.  For the drift
velocity we have $V_d=\langle v_z \rangle$, which is measured in
DNS. The diffusion constant $D_z$ can be estimated by measuring
$\langle T^2 \rangle$ in the DNS and noticing that in
Eq.~(\ref{eq:ig}) $\langle T^2\rangle=L_B(D_z+LV_d/2)/V_d^3$.

Figure~\ref{fig5}b shows the comparison between measured exit-time
PDF $p(T)$ and the inverse Gaussian prediction (\ref{eq:ig}), with
$D_z$ and $V_d$ obtained as discussed above. The prediction results
rather accurate for the right tail (long exit times) for all
turbulent intensities $\gamma$, while the left tail is fairly well
captured only for large values of $\gamma$. The latter deviations can
be interpreted as the result of the suppression of fast escapes due to
gyrotactic trapping, which is more effective in the limit $\gamma \to
0$.  Conversely, long escape times allow trajectories to sum-up many
uncorrelated contributions, thus recovering a diffusive behavior,
which explains the good agreement on the right tail.  The average exit
time of single trajectories, $T_e$, is given by $T_e= L_B/(2\langle
v_z\rangle)$ (as implied by the argument of the exponential in
(\ref{eq:ig}). While for $D_z$ the estimation is more difficult
because the turbulent diffusivity is influenced by swimming,
especially for low $\gamma$ values (see \cite{Santamaria2014} for a
discussion).

The persistence time of the layer $T_p$,
can be heuristically estimated as the time needed for $90\%$ of
the cells to escape the layer, $\int_0^{T_p} p(T) dT \approx 0.9$.
Using (\ref{eq:ig}) one finds that $T_p$  is in the
order of a few (typically $\sim 2-3$) $T_e$ depending on the value of
the vertical diffusivity $D_z$. Ignoring some difficulties in
estimating $D_z$ for realistic oceanic flows, if we consider average
swimming speed $\langle v_z \rangle$ in the range $0.2-0.6 \, v_s$, as
suggested by Fig.~\ref{fig5}a (with $v_s\approx 100-200 \mu
m/s$) and typical lengths $L_B$ of the order of a few centimeters, we
obtain an estimation of $T_p$ from a few hours to a few days, which is
akin to values found in TPLs observed on the field \cite{Sullivan2010}.

\section{Gyrotactic clustering in homogeneous turbulence\label{sec:turbo}}

\begin{figure*}[ht!]
\begin{minipage}[c]{1.1\columnwidth}
\includegraphics[width=\columnwidth]{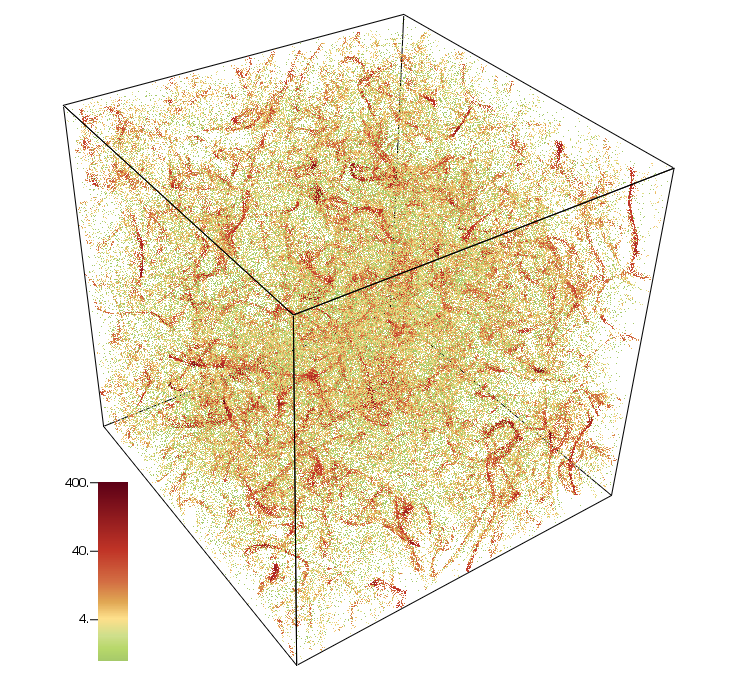}
\put(-250,232){(a)}
\end{minipage}
\begin{minipage}[c]{0.9\columnwidth}
\begin{minipage}[c]{1.0\columnwidth}
\includegraphics[width=\columnwidth]{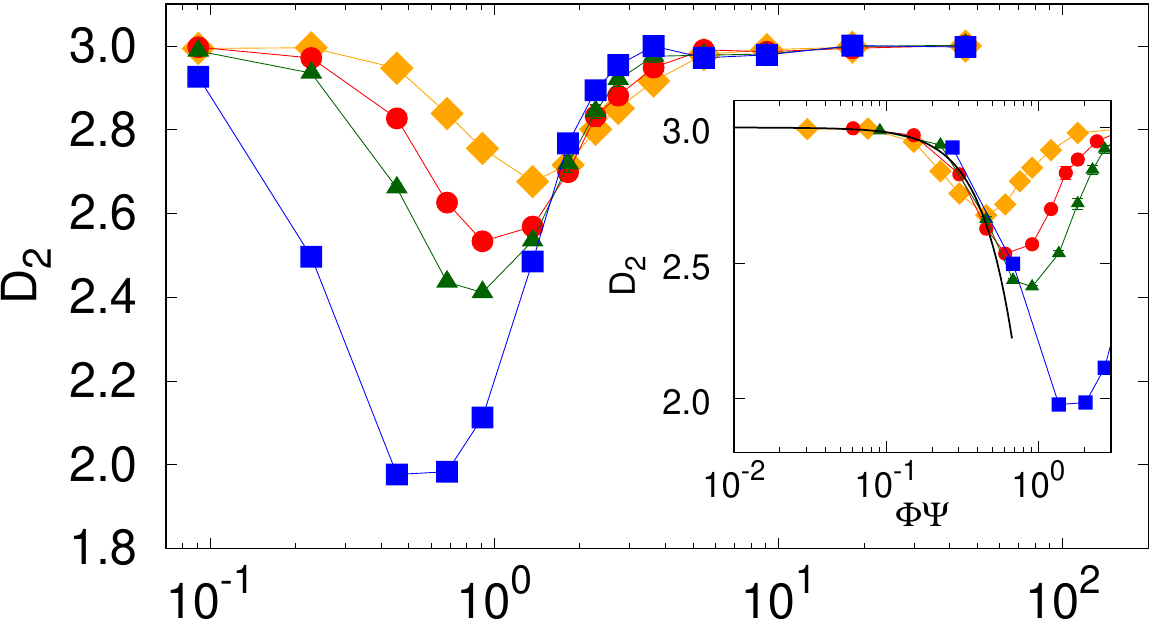}
\put(-215,112){(b)}
\end{minipage}
\begin{minipage}[c]{.985\columnwidth}
\hspace{0.1cm}\includegraphics[width=\columnwidth]{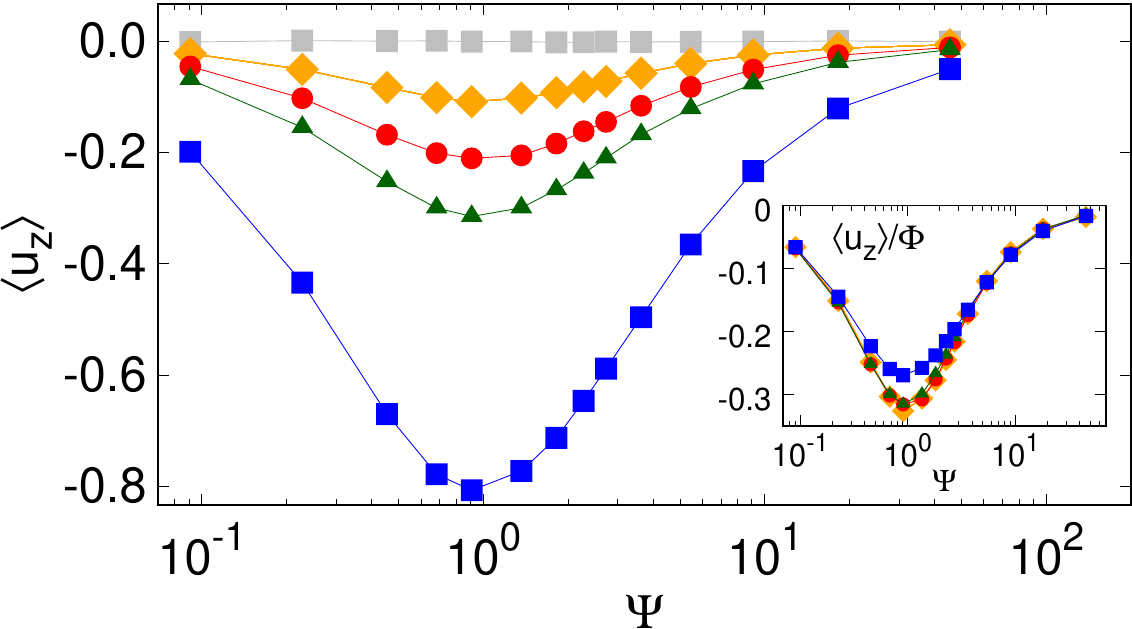}
\put(-215,112){(c)}
\end{minipage}
\end{minipage}
\caption{Clustering and preferential sampling of fluid velocity for
  gyrotactic swimmers in turbulence. (a) Instantaneous configuration
  of $3\times 10^5$ particles, at $\Phi=3$, $\Psi=0.7$. Each particle is colored according to the ratio of the local number density around it to the average density. (b) Correlation dimension $D_2$ of
  cell clusters as a function of the stability parameter for
  non-dimensional swimming speeds $\Phi=1/3$ (yellow diamonds),
  $\Phi=2/3$ (red circles), $\Phi=1$ (green triangles) and $\Phi=3$
  (blue squares). In the inset, $D_2$ vs. $\Phi\Psi$ (left branch of
  the curves shown) with the fitting curve $D_2=3-1.73\Phi^2\Psi^2$,
  compatible with the theoretical prediction of a quadratic
  dependence. (c) The average vertical component of the fluid
  velocity, sampled on particle positions (same symbols as panel
  (b)). Gyrotactic swimmers clearly reside preferentially in
  downwelling regions.}
\label{fig6}
\end{figure*}

As discussed in previous sections, inhomogeneous distributions can arise in dilute suspensions of gyrotactic phytoplankton in laminar flow,
either due to directed motility for stable cells ($\Psi<1$)
\cite{kessler1985hydrodynamic} or due to trapping of unstable cells
($\Psi>1$) in high shear regions \cite{durham2009disruption},
phenomena well reproduced by the mechanistic model defined by
Eqs.~(\ref{eq2.1})-(\ref{eq2.2}). Clustering of cells has been
numerically predicted also in a Taylor-Green vortical flow both for
gyrotactic swimmers \cite{durham2011gyrotaxis} and for elongated,
non-gyrotactic motile cells \cite{Torney2007}.  It is rather difficult
to infer, from these results, the effects of unsteadiness of the flow
or of the presence, typical of turbulence, of many active scales and
intense fluctuations.

While data on large scale phytoplankton patchiness abound
\cite{Martin2003,Mackas1985}, small-scale inhomogeneities have not
been widely studied until recently, when in-situ observations by high
resolution submarine cameras became feasible \cite{Malkiel1999,Gallager2004}.
Although data are still scarce, they provide evidence of clustered distribution
(patchiness) of phytoplankton cells on scales comparable to the dissipative
scales of turbulence \cite{Malkiel1999}. In particular, motile cells exhibit a
more intense patchiness on these scales. One possible interpretation is that
swimming can interact with turbulence to produce inhomogeneous distributions at
small scales.

The above hypothesis seems, apparently, at odds with the mixing
properties of turbulent flow, which typically smooth out
inhomogeneities, e.g., of tracer particles. However, it is a now common
observation in fluid dynamic research that inertial particles,
i.e. finite-size impurities with a density different from that of the
advecting fluid, distribute inhomogeneously in an incompressible flow,
typically in the form of small-scale fractal aggregates. It can be
shown, for example by making use of a standard description of the
force acting on small spherical particles \cite{maxey1983equation},
that particles heavier than the fluid are ejected from vortices, while
lighter particles (such as bubbles) are attracted to the center of
vortices thus forming quasi-one-dimensional clusters in the core of
vortex filaments
\cite{Squires1991preferential,Calzavarini2008dimensionality,bec2007heavy},
a phenomenon dubbed \textit{preferential concentration}.  It is based
on those observations that one can understand how gyrotaxis can indeed
give rise to non-trivial cell distributions in generic turbulent
flows, even without a definite large-scale mean flow.

As we will see in the following, in order for
  small-scale clustering to occur, the swimming speed, $v_s$, and the
  gyrotactic orientation time, $B$, of a cell must be of the order of
  the typical, small-scale speeds and turn-over times of turbulence,
  i.e. of the Kolmogorov velocity $u_\eta\approx (\nu\epsilon)^{1/4}$
  and time $\tau_\eta\approx (\nu/\epsilon)^{1/2}$ ($\nu$ being the
  fluid viscosity) \cite{Frisch1995}. In the ocean, typical turbulence
  intensities, measured in terms of the energy dissipation rate
  $\epsilon$, fall in the range $\epsilon\approx 10^{-8}\div 10^{-4}
  {\rm m^2/s^3}$ \cite{Thorpe2007}.  These values correspond to
  $u_\eta\sim 3 \times 10^{-4}\div 3 \times 10^{-3} {\rm m/s}$ and
  $\tau_\eta\sim 10 \div 10^{-1}{\rm s}$. Given the typical values of
  $B\sim 1\div 10 {\rm s}$ and $v_s\sim 100\div 300 \mu{\rm m/s}$
  \cite{harvey2015persistent,guasto2010oscillatory}, one can see that both
  $\Phi=v_s/u_\eta$ and $\Psi=B/\tau_\eta$ can be of order
  1. 

\subsection{Fractal clustering of gyrotactic phytoplankton in turbulence}

Oceanic turbulence in the bulk of the mixed layer can be considered
as homogeneous and isotropic \cite{Thorpe2007}. Direct numerical
simulations of homogeneous and isotropic turbulent flows seeded with
cells following the dynamics (\ref{eq2.1})-(\ref{eq2.2}), 
relevant for most oceanic applications where fluid accelerations
are much smaller than gravity, showed indeed that intense clustering (as
clear from Fig.~\ref{fig6}a) can occur even in the absence of
coherent large-scale structures or of a mean flow
\cite{durham2013turbulence,zhan2014accumulation}. In analogy to the
case of inertial particles \cite{Bec2003}, this effect can be
explained in the framework of dynamical systems
theory. Equations~(\ref{eq2.1})-(\ref{eq2.2}) describe a dissipative,
potentially chaotic (depending on the advecting flow) system. It has
five, in general non-null, Lyapunov exponents\footnote{Five and not
  six dimensions because ${\bf p}$ is a unit vector.} $\lambda_i$
whose sum gives the rate of change $\dot{\Gamma}$ of an infinitesimal
phase-space volume \cite{ott1993}. The latter is in turn given by the
average divergence in phase space of Eqs.~(\ref{eq2.1})-(\ref{eq2.2}), namely
\begin{equation}
\label{eq5.1}
\dot{\Gamma}=\sum_{i=1}^{3}\left\langle\frac{\partial \dot{x_i}}{\partial x_i}+\frac{\partial \dot{p_i}}{\partial p_i}\right\rangle=-\frac{2}{\Psi} \langle p_z\rangle\,,
\end{equation}
where $B$ has been made non-dimensional in terms of the stability
parameter $\Psi=B/\tau_\eta$.  Though the above average cannot be done
analytically in general, it should be noticed that, in the limit of
small $\Psi$ (and therefore fast orientation), $p_z\to 1$ and the sign
of (\ref{eq5.1}) is negative. This is the signature of the dissipative
nature of the system and it has the consequence that an initially
uniform distribution of cells will converge onto a fractal set
of dimension ${\mathcal D}<5$ in phase space and, consequently, produce
a distribution of particles in physical space with dimension
$D=\min\{{\mathcal D},3\}$ \cite{falconer1986geometry}. If $\mathcal{D}<3$
the dimension in real space will be $D=\mathcal{D}<3$ signalling
the presence of observable clustering, indeed
a homogeneous distribution would give $D=3$.

Figure~\ref{fig6}b shows the fractal dimension $D$ as a function of
$\Psi$ for different values of the swimming parameter $\Phi$. The
particular definition of $D$ used here is the correlation dimension
$D_2$, which represents the scaling exponent of the $r$-dependence of
the probability $p_2(r)\sim r^{D_2}$ to find a pair of particles with
separation less than $r$ \cite{grassberger1983characterization}. This
is, of course, only one possible definition of $D$ as cell
distributions will in general be multifractal
\cite{paladin1987anomalous}. The qualitative behavior of $D_2$ as a
function of the stability parameter $\Psi$ can be understood
considering the limits of fast and slow orientation. In the limit
$\Psi\to 0$, cells rapidly orient upwards, so that one can write
$\dot{\bf x}={\bf u}({\bf x},t)+\Phi \hat{\bf z}$, in non-dimensional
units. The latter expression has zero divergence and leads to no
accumulation, the dynamics collapses onto the real space, where it
preserves the volume \cite{ott1993}. In the opposite limit $\Psi\to
\infty$ cells are unstable and randomly tumble due to the local
vorticity fluctuations. Moreover, in this limit, as clear from
Eq.~(\ref{eq5.1}) $\dot{\Gamma}\to 0$, meaning that the dynamics
preserve phase-space volumes so that one expects $\mathcal{D}\to 5$ and
thus $D\to 3$. Again this means no clustering. Any clustering must
therefore happen for intermediate values of $\Psi$, as indeed shown in
Fig.~\ref{fig6}b.

  The above phenomenological argument suggests the
  possibility to obtain a quantitative theory for clustering in the
  limit of fast orientation ($\Psi\ll 1$). In order to make the
  derivation clear it is useful to better define the parameter range
  in which we are working: to this aim we refer to the theory
  developed in Ref.~\cite{fouxon2015phytoplankton}. First of all we
  notice that the orientation dynamics (\ref{eq2.1}) is fully
  determined by the history of the velocity gradients along the
  trajectory of the particle. 
  In turbulence the typical correlation
  time of velocity gradients along a tracer path is of the order of
  $\tau_\eta$. However, for swimming particles such correlation time
  depends on the swimming speed. Indeed when $\Phi>1$,
  i.e. $v_s>u_\eta$, the correlation time will be of the order of
  $\eta/v_s=\tau_\eta/\Phi$, being the Kolmogorov length $\eta$ the
  typical scale of the gradients. Hence the correlation time of the
  vorticity along the swimming particle path will be
  $t_{cor}=\min\{\tau_\eta,\tau_\eta/\Phi\}$. As detailed in
  Ref.~\cite{fouxon2015phytoplankton}, from Eq.~(\ref{eq2.1}), the
  above observation and the fact that the vorticity magnitude is order
  $1/\tau_\eta$ one can derive that ${\bf p}$ is weakly affected by
  turbulence under two circumstances: either when $\Psi \ll
  t_{cor}/\tau_\eta=\min\{1,1/\Phi\}$, or when $\Psi\gg
  t_{cor}/\tau_\eta$ and $\Psi/\Phi\ll 1$. The latter regime
  corresponds to the case in which the gradients seen by the particle
  are almost uncorrelated and the vertical polarization results from
  the central limit theorem. In this case precise predictions on the
  PDF of the orientation can be obtained
  \cite{fouxon2015phytoplankton}. In the former regime, instead, the
  vorticity changes slowly with respect to the orientation dynamics,
  and local equilibrium is a good approximation. Therefore,
  similarly to what was done in Sect.~\ref{sec2} to explain cell focusing, one can
  impose $\dot{\bf p}=0$ and expand (\ref{eq2.1}) to first order in
  $\Psi$ obtaining the following expression for the instantaneous
  equilibrium orientation,
\begin{equation}
\label{eq5.2}
{\bf p}_{eq}\sim\left(\Psi\omega_y,-\Psi\omega_x,1\right)\,.
\end{equation}
In Ref.~\cite{fouxon2015phytoplankton} the reader can find a detailed
discussion of the limit of validity of the flow description
(\ref{eq5.2}) in terms of the appearence of singularities in the
gradients of the particle orientation, which are analogous to caustics in inertial
particles \cite{wilkinson2005caustics}.
We now discuss the consequence of Eq.~(\ref{eq5.2}).

First, in this limit the equations of motion reduce to
$\dot{\bf x}={\bf v}={\bf u}({\bf x},t)+\Phi {\bf p}_{eq}$. Thus, the swimmers behave
as tracers transported by an effective velocity field $\bf v$ with divergence
\begin{equation} 
\label{eq5.3}
\nabla\cdot{\bf v}=\Phi \bm \nabla \cdot {\bf p}_{eq}=-\Psi\Phi\nabla^2u_z,
\end{equation}
 Thus the effective velocity field is
 the sum of an incompressible part (the fluid velocity) and a
 compressible one proportional to $\Phi\Psi$.  This situation is in
 close analogy with that found when considering the
 small-Stokes-number limit for inertial particles
 \cite{Falkovich2002,Balkovsky2001,Fouxon2012}.  Following closely those
 derivations, one can show that $3-D\sim (\Phi\Psi)^2$ as indeed
 observed in the inset of Fig.~\ref{fig6}b, where the left branches of
 the curves approximately rescale when plotted as a function of
 $\Phi\Psi$ and are well approximated by a parabola.
 It is worth observing that the $(\Phi\Psi)^2$
   dependence for the codimension of the fractal clusters is based on
   (\ref{eq5.3}) and is valid when both $\Psi \ll 1$ and $\Phi\ll
   1$. However, it was then shown \cite{fouxon2015phytoplankton} (see
   also \cite{gustavsson2016preferential}) that the same dependence
   should be expected also when $\Phi\gg 1$ provided $\Psi \ll
   1/\Phi$, though with a different prefactor. These derivations are
   based on an estimation of the Kaplan-Yorke dimension
   \cite{fouxon2015phytoplankton} and perturbative results based on a 
   small-Kubo expansion \cite{gustavsson2016preferential}. The analogy with weakly-compressible flows allows one to extend this argument to the generalized fractal co-dimension of any order $q$, which is predicted to depend linearly on $q$ \cite{Fouxon2012}.

 Secondly, trajectories will tend to concentrate in regions where the
 divergence of the effective flow is negative. Therefore, from the
 expression for $\dot{\Gamma}$ we can predict that cells will
 preferentially sample regions where $\nabla^2u_z>0$.  By
 considerations of isotropy of the velocity field and a positive
energy dissipation at small scales, one can conclude
 that such regions are also, on average, regions where $u_z<0$
 \cite{durham2013turbulence}.
 In other words, fast re-orienting
 gyrotactic swimmers preferentially sample downwelling regions of the
 flow, even when the background flow is turbulent and statistically
 isotropic.\footnote{Clearly, the symmetry breaking happens at the
   level of the dynamics (\ref{eq2.1}).} This conclusion is confirmed
 in Fig.~\ref{fig6}c, showing that the average vertical fluid speeds
 at cells' positions is negative and, moreover, as shown in the inset
 rescale with the swimming parameter $\Phi$.

 The above argument to explain the prefential
   sampling of downwelling flow is based on an Eulerian average and
   not on the average on the particle position. In
   \cite{fouxon2015phytoplankton} a different, more
   precise, derivation was obtained based on the observation that
   denoting with $n({\bf x},t)$ the density of swimming particles, the
   average flow vertical velocity at the particle positions is
   $\langle u_zn\rangle$. Then one observes that in the limit in which
   they can be described as tracers in a compressible flow, $n$
   evolves as $\dot{n}=-\bm \nabla\cdot{\bf v}=\Psi\Phi\nabla^2 u_z$,
   where (\ref{eq5.3}) is used. By the formal solution of the above equation
   one can write
   \begin{equation}
     \langle u_zn\rangle=  \Psi\Phi\int_{-\infty}^{t} ds
     \langle u_z({\bf x}(t),t) \nabla^2 u_z({\bf x}(s),s)\rangle\,.
\label{eq:fouxon}
   \end{equation}
Again, to show that $\langle u_zn\rangle<0$, the Authors of
 \cite{fouxon2015phytoplankton} assume that, at equal time, $\langle
 u_z\nabla^2 u_z\rangle=-\langle (\nabla u_z)^2\rangle<0$. Moreover,
 in the regime in which $\Phi\gg 1$ and $\Psi\ll 1/\Phi$ the negative
 sign can be proved computing explicitly the correlation in
 (\ref{eq:fouxon}) as an Eulerian correlation, because, thanks to the
 fast swimming speed, the turbulent velocity field can be assumed as
 frozen on the time scales involved. We also mention that the
 preferential sampling of downwelling regions was also derived,
 following a completely different approach, in
 Ref.~\cite{gustavsson2016preferential}, where sharp analytical
 results were obtained for random flows.

\begin{figure}[t!]
\includegraphics[width=1.0\columnwidth]{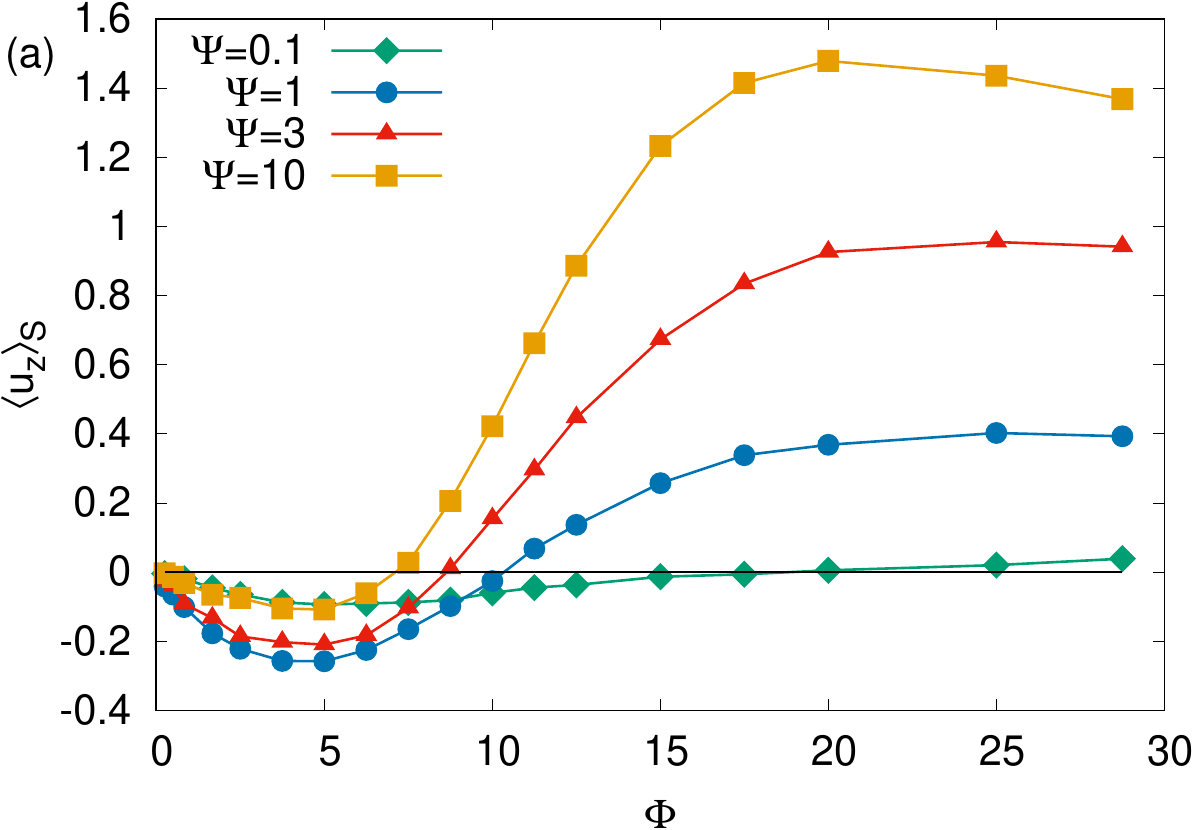}
\includegraphics[width=1.0\columnwidth]{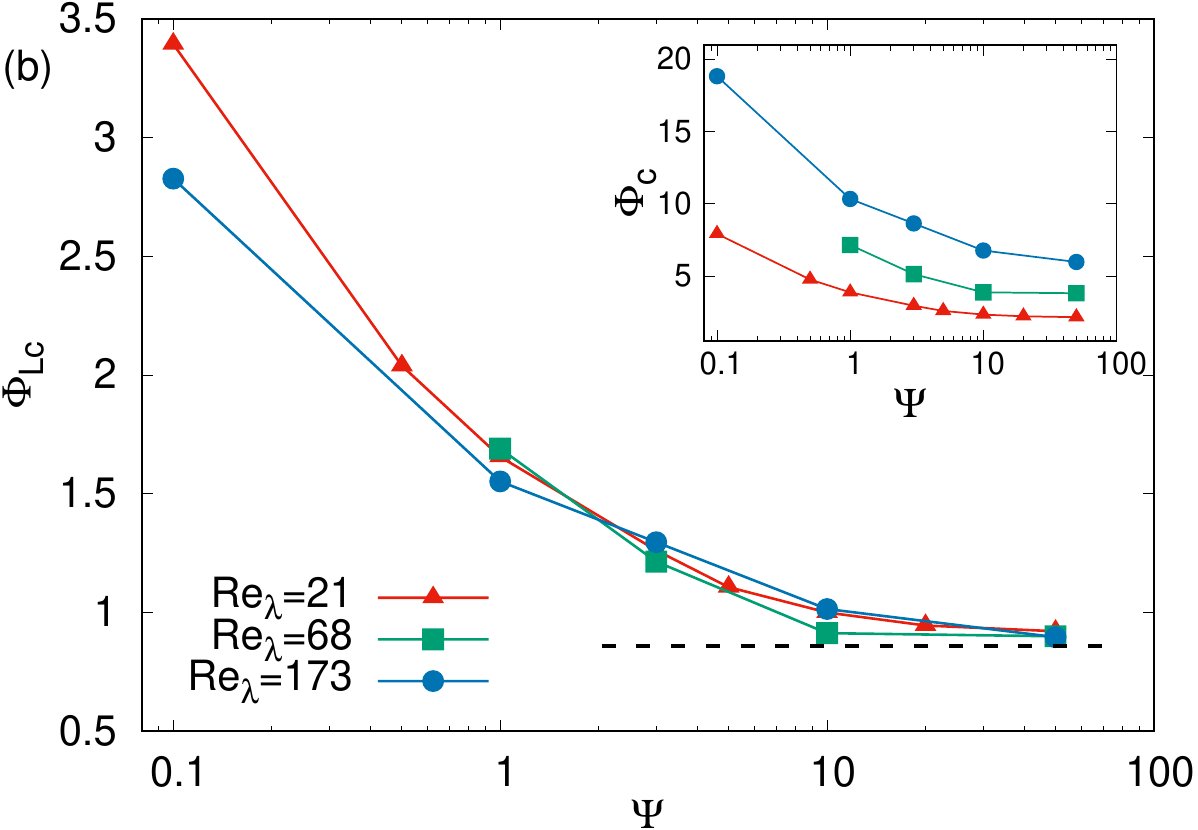}
\caption{Effects of shape on the preferential sampling
    of fluid velocity. (a) Average of the vertical component of fluid
    velocity conditioned on swimmer position, computed along the
    trajectory of rod-like gyrotactic particles. For swimming
    parameters larger than a ($\Psi$-dependent) critical value, the
    swimers preferentially sample up-welling regions, at variance with
    spherical particles. (b) While the critical value $\Phi_C$ for the
    inversion of $\langle u_z\rangle$ depends on Re (inset), the same
    cuves collapse together when plotted in terms of large-scale
    parameter $\Phi_{L,C}$ based on $u_{rms}$ (main plot, see text).
}
\label{fig7}
\end{figure}

In Refs.~\cite{zhan2014accumulation} the model (\ref{eq2.1}) was studied, 
by means of numerical simulations, for generic values of $\alpha$ thus taking
into account the effect of the cell's shape. 
Theoretical predictions as a function of $\alpha$ were obtained
in Refs.~\cite{gustavsson2016preferential}, some of which were recently
numerically confirmed in turbulent flows Refs.~\cite{borgnino2018gyrotactic}.
For elongated, ellipsoidal cells, besides
vorticity also the strain rate can indeed influence the cell orientation.
Remarkably, it was found, both by theoretical considerations
\cite{gustavsson2016preferential} and by numerical simulations
\cite{gustavsson2016preferential,borgnino2018gyrotactic} that when the cells'
aspect ratio increases, i.e. they become more rod-like, and for large swimming
velocities the preferential sampling reverses and cells spend more time in
upwelling regions. Here we will focus for simplicity on the case of rods. The
effect can be appreciated in Fig.~\ref{fig7}(a), showing the average of $u_z$
computed on the positions of rod-like swimmers in a DNS of homogeneous
turbulence \cite{borgnino2018gyrotactic}. Clearly there is a critical swimming
speed $\Phi_c(\Psi)$ above which they experience positive, instead of negative,
average vertical fluid speeds. If this analysis is repeated at varying Re, the
observed value of $\Phi_C$ increases with Re (Fig.~\ref{fig7}(b), inset) and
appears to scale with $u_{rms}$ (which is a large scale quantity), as can be
seen in Fig.~\ref{fig7}(b)  (main panel) where a large-scale swimming
parameter $\Phi_L=v_s/u_{rms}$ has been introduced. The latter results lead one
to conclude that the inversion in the preferential sampling is controlled by
velocity correlations, while the small-$\Phi$ dinamics is controlled by
velocity-gradient correlations. In developed turbulence, the two quantities
have different characteristic times, namely $\tau_\eta$ for the gradients and
the large-eddy turnover time $\tau_L$ for velocities, with
$\tau_L/\tau_\eta\sim {\rm Re}^{1/2}$, so that the separation of the two
dynamics increases at increasing Re.  Differences between the elongated and the
spherical case were found also for fractal clustering, both in stochastic
models \cite{gustavsson2016preferential} and in DNS of homogeneous turbulence
\cite{zhan2014accumulation,borgnino2018gyrotactic} It was found that the
intensity of clustering, i.e.  the extent of the deviation of $D_2$ from the
homogeneous value $3$, decreases for such more elongated cells.  However,
elongated cells display a tendency to accumulate even in the absence of
gyrotaxis ($\Psi\rightarrow\infty$). A theoretical framework for this phenomenon
was provided in the context of stochastic models \cite{gustavsson2016preferential}.

\subsection{Effects of turbulent accelerations on gyrotactic clustering}
\begin{figure}[t!]
\includegraphics[width=1.0\columnwidth]{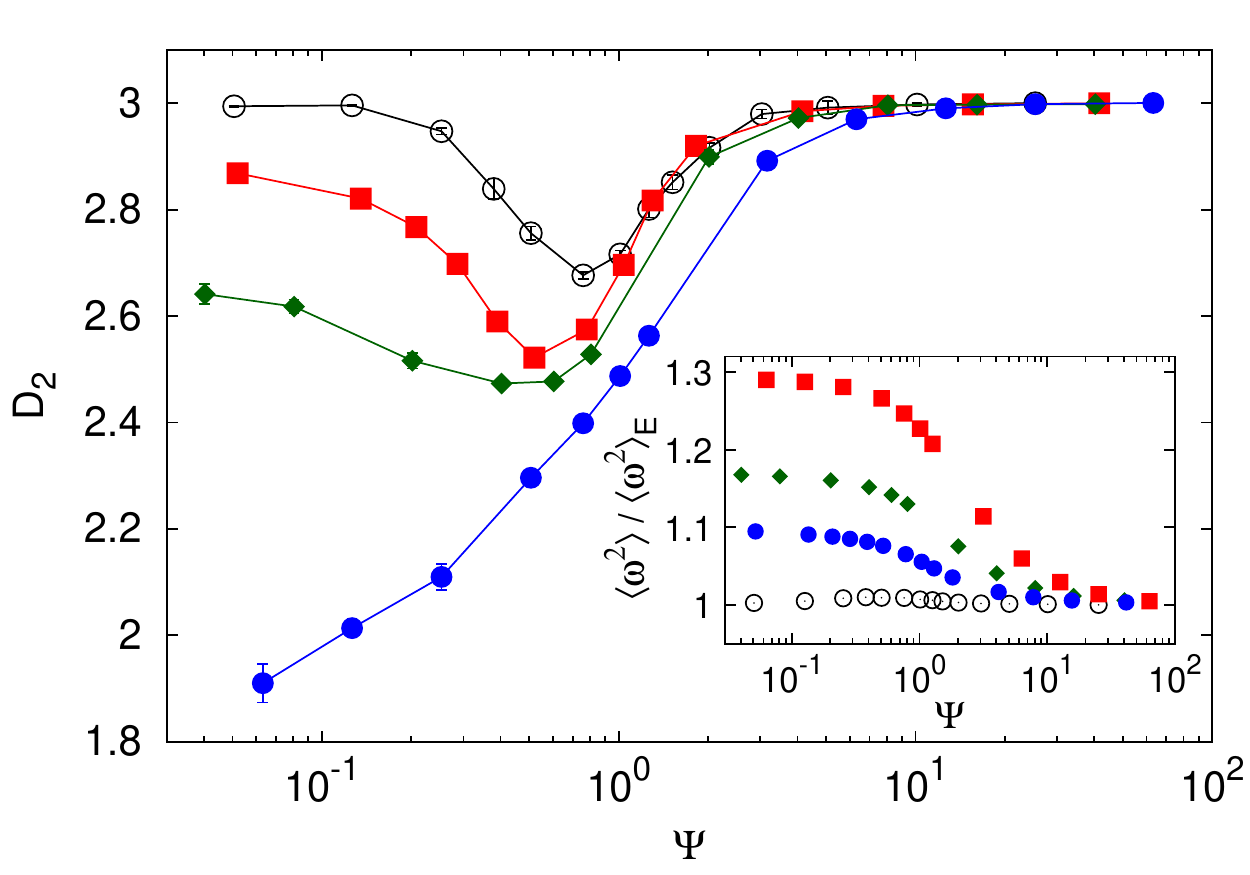}
\caption{Clustering and preferential sampling of high vorticity
  regions when fluid acceleration is considered. Main plot:
  correlation dimension of swimmer clusters. Inset: mean square
  vorticity computed over particle positions and normalized with the
  Eulerian average. In all simulations $\Phi=1/3$ both for the
  model considering only gravity, i.e. eq.~(\ref{eq2.1}) at $Re_\lambda=62$ (empty circles), and including the effect of fluid acceleration
  (eq.~(\ref{eq2.3}), filled symbols) at $Re_\lambda=20$ (squares),
  $36$ (diamonds), $62$ (circles).}
\label{fig8}
\end{figure}

In the case of intense turbulence, swimmers may experience extreme
accelerations \cite{LaPorta2001}, which in principle require the use
of model~(\ref{eq2.3}) instead of (\ref{eq2.2}).  As seen from the
experimental results in Fig.~\ref{fig1}, when centrifugal
acceleration exceeds gravity swimmers are expected to concentrate
towards the center of a stationary vortex. Again one can wonder what
would happen when unsteady, turbulent flows are considered.  Direct
numerical simulations model (\ref{eq2.1})-(\ref{eq2.3}) in homogeneous
and isotropic turbulence \cite{delillo2014turbulent} show that indeed
this dynamics is relevant also in more general flows.
Figure~\ref{fig8} shows the correlation dimension of cell clusters
in the case where fluid acceleration is explicitly
considered. Comparing with Fig.~\ref{fig6}b, the main qualitative
change due to fluid acceleration is observed for fast orientations
($\Psi\ll 1$): at increasing turbulence, and thus the weight of fluid
acceleration, $D_2$ deviates more and more from the uniform value $3$
also when $\Psi\to 0$. In the limit of negligible gravity, $D_2$ is
monotonous in $\Psi$ and one has maximum clustering at $\Psi\rightarrow
0$.

When fluid acceleration is very intense and gravity negligible, the
dynamics is better described in terms of a stability number based on
the typical acceleration encountered $\Psi_a=\omega_{rms}v_0/a_{rms}$.
With the same ideas used in the previous case, it is easy to realize
that when $\Psi_a$ is small, cells will orient along the local
acceleration so that an effective velocity field can be written as
${\bf v}=u+\Phi\hat{\bf a}$ (with $\hat{\bf a}={\bf a}/|{\bf a}|$).
Again one has that swimmers behave as tracers in a compressible velocity
field as $\bm \nabla \cdot \hat{\bf a}\neq 0$ and they will
concentrate in regions characterized by $\bm \nabla\cdot\hat{\bf
  a}<0$. Numerical simulations lead to conclude that the signs of
$\bm \nabla\cdot\hat{\bf a}$ and $\bm \nabla\cdot{\bf a}$ are strongly
correlated. The request that the latter quantity be negative is
tantamount to requiring that the flow is locally dominated by
vorticity, since
\begin{equation}
\bm \nabla\cdot{\bf a}=\sum_{ij}\left(\hat{S}^2_{ij}-\hat{\Omega}^2_{ij}\right),
\end{equation}
with $\hat{S}_{ij}=\frac12 (\sigma_{ij}+\sigma_{ji})$ and
$\hat{\Omega}_{ij}=\frac12 (\sigma_{ij}-\sigma_{ji})$ being the
symmetric (rate of strain) and antisymmetric (vorticity) part of the
velocity gradient $\sigma_{ij}=\partial_ju_i$, respectively. At
increasing $Re_\lambda$, therefore, gyrotactic swimmers are more and
more attracted towards regions of high vorticity. This is confirmed in
the inset of Fig.\ref{fig8}, showing the average mean square vorticity
measured at the position of cells is plotted. As one can see, swimmers
experience larger vorticities than tracers (which would remain
homogeneous and therefore be subject to the Eulerian average
$\langle\omega^2\rangle_E$) or purely gravitactic swimmers, which also
show very small deviations from the Eulerian value. Remarkably, the
above argument is essentially the same one can use to explain the
clustering of light inertial particles in turbulent vortex filaments
\cite{Balkovsky2001,Calzavarini2008dimensionality}.

\subsection{Clustering of polydisperse populations}

In the above discussion, fractal clustering has been numerically
demonstrated for monodisperse suspensions, i.e. when all cells have
identical stability and swimming parameters. 
Natural colonies are of course characterized by 
a distribution of $\Phi$ and $\Psi$,
it is thus natural to wonder whether fractal clustering can be observed
in realistic conditions.
We discuss here the robustness of clustering in two cases: when the
swimming parameters are Gaussian distributed within the swimmers
population and when two sub-populations are present with different parameters,
e.g. representing two hypothetical strains.
These two cases will be referred as Gaussian and bimodal cases.

When considering two sub-population of swimmers, we necessarily have
to extend the previous definition of $D_{2}$ to quantify a
\textit{cross correlation dimension} \cite{Bec2005,borgnino2017scale};
the latter can be defined via the probability of finding two swimmers,
characterized by two values of parameters $(\Phi_{1},\Psi_{1})$ and
$(\Phi_{2},\Psi_{2})$, at a distance smaller than $r$, namely
$P_{12}(r) \propto r^{D^{(12)}_2}$.  Clearly, for a monodisperse
population (i.e.  $\Phi_{1}=\Phi_{2}$ and $\Psi_{1}=\Psi_{2}$), one
recovers the correlation dimension introduced in the previous section.
In order to understand what we should expect for such quantity, it is
useful, for simplicity, to consider the case in which the
two sub-populations are characterized by the same swimming number
$\Phi=\Phi_1=\Phi_2$ and a small mismatch in the value of the
stability parameter $\Delta\Psi=\Psi_{1}-\Psi_{2}$.  Using
non-dimensional units, from Eq.(\ref{eq2.2}) one can write that the
separation ${\bf R}$ between two swimmers evolves according to
$\dot{\bf R}=\Delta{\bf u}({\bf R})+\Phi \Delta{\bf p}$. At very small
separation, i.e. below the Kolmogorov length $\eta$, the velocity
field is smooth and we can write $\Delta{\bf u(R)}\sim u_\eta (R/
\eta)$.  Thus from the balance of the two terms in the above equation
a characteristic scale emerges \cite{borgnino2017scale}:
\begin{equation}
R^{*}\simeq \eta \Phi \Delta\Psi\,.
\end{equation}
A similar argument can be done for swimmers with the same stability
parameter and different swimming velocities where one finds
$R^{*}\simeq\eta\Psi\Delta\Phi$ \cite{borgnino2017scale}.  For
$R<R^{*}$ the separation dynamics is dominated by the parameter
mismatch and one expect poor correlation between the swimmers, i.e.
homogeneous distribution.

\begin{figure}[b!]
\includegraphics[width=1.0\columnwidth]{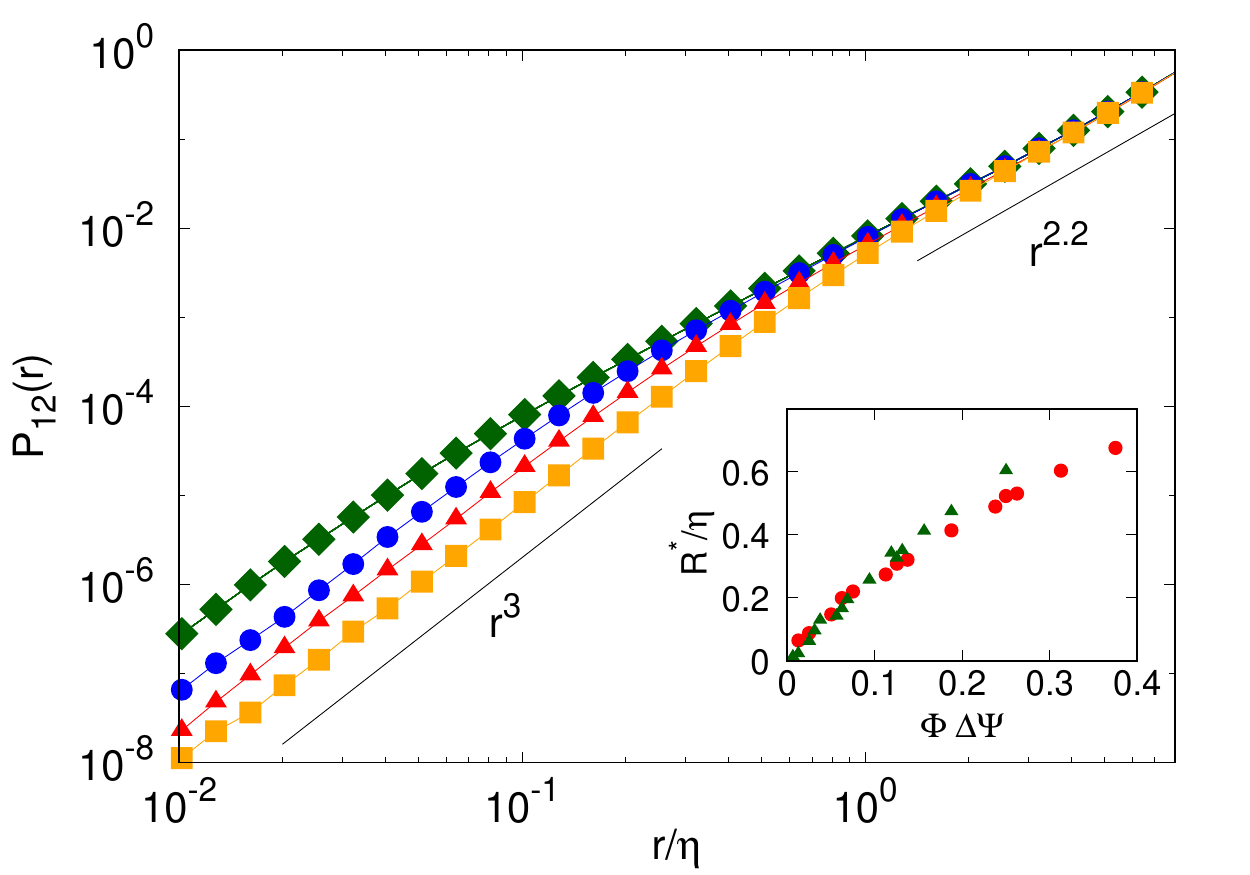}
\caption{Probability $P_{12}(r)$ for different pairs of sub-populations
  characterized by constant $\overline{\Psi}=0.57$ and by $\Delta\Psi =0.0042$ (green diamonds),
  $\Delta\Psi =0.021$ (blue circles), $\Delta\Psi =0.042$ (red
  triangles) and $\Delta\Psi=0.125$ (orange squares).
  Note that $\Phi=3$ for all curves.
  (Inset) Crossover scale as a function of $\Phi\Delta\Psi$ for two different
  sets of populations with $\Phi=3$ (red circles) and $\Phi=1.5$
  (green triangles).}
\label{fig9}
\end{figure}

For $R>R^{*}$, conversely, the dynamics is dominated by the velocity
field and it is almost indistinguishable from that of two cells with
the same stability parameter. In this latter case, one expects to
observe fractal clustering. In practice, this means that the
cross correlation dimension will depend on $r$ and
$D_2^{(12)}(r)\approx 3$ for $r<R^{*}$ and $D_{12}(r)<3$, signaling
fractal clustering, above $R^*$. This is very similar to what happens
to inertial particles with slightly different Stokes times
\cite{Bec2005}. This scenario is confirmed in Fig.~\ref{fig9}, which
shows the cross probability $P_{12}(r)$ that quantifies the
case of two sub-population with different values of
$\Psi$. Curves obtained with different $\Delta \Psi$ but the
same $\overline{\Psi}=\frac12(\Psi_1+\Psi_2)$ display the same
behavior at very small scale, where the scaling exponent is close to
$3$, thus the sub-populations considered see each other as uniformly
distributed or, in other words, they have uncorrelated distributions.
Conversely, at larger scales, a nontrivial power-law behaviour is
observed with a exponent close to that one of a homogeneous population
with stability number $\overline{\Psi}$. Finally, the inset of
Fig.~\ref{fig9} confirms the linear scaling of the characteristic
scale $R^{*}$ discussed above by showing a remarkable collapse of the
different crossover scales when plotted as a function of the
combination $\Phi\Delta\Psi$. Similar results, both concerning the
cross probability and the characteristic scale, can be obtained when a
bimodal distribution with two different swimming numbers $\Phi_{1}$
and $\Phi_{2}=\Phi_{1} - \Delta\Phi$ and same $\Psi$ is considered.

\begin{figure}[h!]
\includegraphics[width=1.0\columnwidth]{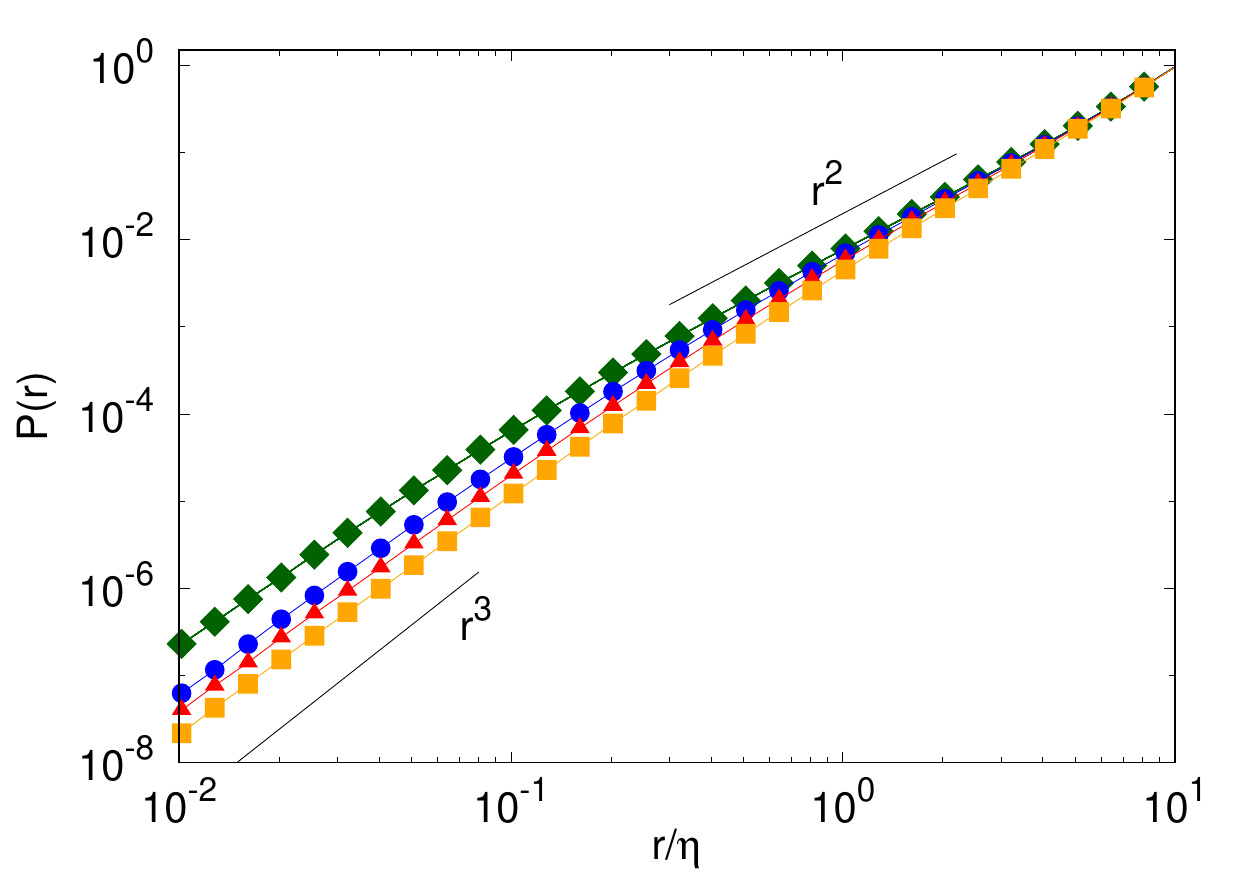}
\caption{Probability $P(r)$ for a population of swimmers with fixed
  $\Phi=3$ and $\Psi$ Gaussian distributed with
  $\overline{\Psi}=0.583$ and $\sigma_{\Psi}=0.008$ (green diamonds),
  $\sigma_{\Psi}=0.042$ (blue circles), $\sigma_{\Psi}= 0.083$ (red
  triangles) and $\sigma_{\Psi}= 0.166$ (orange squares).}
\label{fig10}
\end{figure}

We now consider a more realistic polydisperse suspension, whose stability
number $\Psi$ (or equivalently $\Phi$) is Gaussian distributed with
mean value $\overline{\Psi}$ and standard deviation
$\sigma_{\Psi}$. In particular, assuming a $\sim30\%$ of relative
variation in gyrotactic parameters, which is rather realistic
\cite{sengupta2017phytoplankton,harvey2015persistent}, is fractal
clustering still observable?  Here, it is useful to consider the
cumulative probability, $P(r)$, of having two cells separated by a
distance less than $r$ integrated on all the possible pairs within the
population.  In the same way of the bimodal distribution, as shown in
Fig.~\ref{fig10}, $P(r)$ is characterized by two different scaling
regions: at very small scales all curves recover a uniform scaling
$r^{3}$, while for $r \gtrsim\eta$ fractal clustering can be observed,
as signaled by a power-law behavior of $P(r)$ with a scaling exponent
very close to that expected for a monodisperse population with
$D_{2}=D_{2}(\overline{\Psi})\simeq2$, but for a weak dependence on
$\sigma_{\Psi}$. The exponent for the smallest $\sigma_\Psi$ is
$\simeq 2$ while it grows to above $2.3$ for the population with the
largest value of $\sigma_\Psi$.  Also, similarly with the bimodal
case, the transition between the two different scaling ranges shifts
to larger scales as $\sigma_{\Psi}$ is increased.  Remarkably, also
the case with largest standard deviation
($\sigma_{\Psi}/\overline{\Psi}\simeq0.29$) shows a strongly
inhomogeneous distribution at large scale indicating the presence of
fractal clustering even in a population with a high variability in the
gyrotactic parameters. This result provides a strong indication that
fractal clustering should be detectable also in experimental
suspensions.

\section{Conclusions and perspectives \label{sec:conclusions}}
The range of flow effects on motile microorganisms here reviewed is by
no means exhaustive. We mostly focused on those aspects which can be
well characterized and understood by using tools and ideas from
dynamical systems. To this aim we have specialized our discussion on
gyrotactic motility showing that ideas from integrable (conservative)
dynamical systems can explain the phenomenon of gyrotactic trapping in
terms of barriers to transport induced by the separatrices emerging
from hyperbolic fixed point of the dynamics. Similar ideas can, and
have been used, e.g., to explain the behavior of bacterial
trajectories in shear flows \cite{Stark_PRL2012,Stark_EPJE2013}.  On
the other hand the physics of dissipative, chaotic dynamical systems
has been used to successfully explain micro-patchiness of gyrotactic
phytoplankton in turbulent flows. Again similar ideas can be extended
to other kind of motile microorganisms \cite{Torney2007} and also to
artificial micro-swimmers, such as phoretic colloids
\cite{schmidt2016clustering,shukla2017phoresis}.

More in general we think that the kind of approach here reviewed can
be helpful, possibly by also accounting for stochastic effects (which
can also alter the swimming direction via, e.g., rotational Brownian
motion or intrinsic stochasticity of the propulsion mechanisms) for
understanding most of the effects arising from the interplay between
flow and motility in dilute suspensions of living or artificial
micro-swimmers. In particular, we believe there are many directions
that still need to be explored or fully understood.

For instance, the dynamics of nonspherical gyrotactic organisms 
\cite{zhan2014accumulation,gustavsson2016preferential,borgnino2018gyrotactic} certainly deserves further study in view of its richer phenomenology.
Still in the context of gyrotactic motility, some recent experiments
have shown that gyrotactic cells can, to some extent, display some
level of adaptation when exposed to frequent reorientation, such as
those induced by turbulence, for a long time
\cite{sengupta2017phytoplankton}. It would thus be very interesting to
model and characterize such behaviors. Also the motility of other kinds
of phytoplankton cells have been shown to be influenced by shear
flows, for instance, at high shear rate, \textit{Dunaliella
  primolecta} swims in the direction of local flow vorticity
\cite{chengala2013microalga}. Interestingly, this behavior seems to be
the result of active shear-induced response and this opens several
interesting questions both in the direction of understanding the origin
of such adaptation and the consequences for their spatial
distribution.

Another direction of interest is to investigate the combined effect of
flow and directed motility as due to chemotaxis. As briefly discussed
in the introduction, the presence of shear can trap bacteria and
deplete their chemotactic efficiency \cite{rusconi2014bacterial}. This
effect is mostly due to Jeffery orbits which align elongated cells
along the shear. How and to what extent similar phenomena alter the
motility in unsteady, turbulent flows, such as those encountered in
the oceans, is largely unknown. In the presence of turbulence, also
the chemical field is advected and mixed by the flow. What is the
effect on chemotaxis? Some attempt in this direction
\cite{taylor2012trade} has shown that the chemotactic response can be
altered by turbulence also without considering the direct impact of
the flow on motility. It would thus be very interesting to include and
quantify such effects. This is particularly interesting in view of the
fact that aquatic bacteria often display different motility strategies
with respect to enteric ones \cite{stocker2012ecology}, which may be
an indication of adaptation to the exposure to turbulent motion for
such microbes.

\begin{acknowledgments}
We thank M. Barry, E. Climent, W. M. Durham, M. Franchino F. Santamaria, R. Stocker, for their contribution to the papers that form the basis of this review.
We acknowledge support from the COST Action MP1305
``Flowing Matter''. G.B., M.B. and F.D. acknowledge support from the ``Departments of Excellence'' (L. 232/2016) grant,
funded by the Italian Ministry of Education, University and Research (MIUR), Italy. 
\end{acknowledgments}

\section*{Authors contribution statement}
All the authors contributed equally to the writing of this review.


\end{document}